\documentclass[aps,prb,superscriptaddress,floatfix,,preprint,letterpaper]{revtex4-1}
\pdfoutput=1

\newcommand{\ml}[1]{\(#1\)}
\newcommand{\ma}[1]{\begin{align}#1\end{align}}

\begin{document}

\begin{titlepage}

\title{
Pattern-of-zeros approach to Fractional quantum Hall states
and\\
a classification of symmetric polynomial of infinite variables 
}

\author{Xiao-Gang Wen}
\affiliation{Department of Physics, Massachusetts Institute of
Technology, Cambridge, Massachusetts 02139, USA}
\affiliation{Institute for Advanced Study, Tsinghua University,
Beijing, 100084, P. R. China}

\author{Zhenghan Wang}
\affiliation{
Microsoft Station Q, CNSI Bldg. Rm 2237, University
of California, Santa Barbara, CA 93106 }

\begin{abstract}
Some purely chiral fractional quantum Hall states are described by symmetric or
anti-symmetric polynomials of infinite variables.  In this article, we review a
systematic construction and classification of those fractional quantum Hall
states and the corresponding polynomials of infinite variables, using the
pattern-of-zeros approach.  We discuss how to use patterns of zeros to label
different fractional quantum Hall states and the corresponding polynomials.  We
also discuss how to calculate various universal properties (\ie the quantum
topological invariants) from the pattern of zeros.

\end{abstract}


\maketitle

\vspace{2mm}

\end{titlepage}

{\small \setcounter{tocdepth}{1} \tableofcontents }

\vfill
\break

\section{Introduction}

To readers who are interested in physics, this is a review article on the
pattern-of-zeros approach to fractional quantum Hall (FQH) states. To readers
who are interested in mathematics, this is an attempt to classify symmetric
polynomials of infinite variables and $Z_n$ vertex algebra.  To those
interested in mathematical physics, this article tries to provide a way to
systematically study pure chiral topological quantum field theories that can be
realized by interacting bosons.  In the next two subsections, we will review
briefly the definition of quantum many-boson systems, and the definition of
quantum phase for non-physicists.  Then, we will give an introduction of the
problems studied in this paper.

\subsection{What is a quantum many-boson system}

The fermionic FQH states\cite{TSG8259,L8395} are described by anti-symmetric
wave functions, while the bosonic FQH states are described by symmetric wave
functions.  Since there is an one-to-one correspondence between the
anti-symmetric wave functions and the symmetric wave functions, in this
article, we will only discuss bosonic FQH states and their symmetric wave
functions.

Bosonic FQH systems are quantum many-boson systems.  Let us first define
mathematically what is a quantum many-boson system, using an $N$-boson system
in two spatial dimensions as an example.  A many-body state of $N$ bosons is a
symmetric complex function of $N$ variables 
\begin{align}
& \ \ \ \Psi( \v r_1,...,\v r_i,...,\v r_j,...,\v r_N) 
\nonumber\\
&=
 \Psi( \v r_1,...,\v r_j,...,\v r_i,...,\v r_N) 
\end{align}
where the $i^\text{th}$ variable $\v r_i=(x_i,y_i)$ describes the coordinates of
the $i^\text{th}$ boson.  All such symmetric functions form a Hilbert space
where the normal is defined as
\begin{align}
\<\Psi|\Psi\>=
\int \prod_i \dd x_i \dd y_i \;
\Psi^*\Psi
\end{align}

A quantum system of $N$ bosons is described by a
 Hamiltonian, which is a Hermitian operator
in the above Hilbert space. It may have a form
\ma{
H(g_1,g_2)= \sum_{i=1}^N
-\frac12 ( \prt_{x_i}^2 +\prt_{y_i}^2 ) +\sum_{i<j}
V_{g_1,g_2}(\v r_i-\v r_j)
}
Here $V_{g_1,g_2}(\v r_i-\v r_j)$ is the interaction potential between two
bosons. We require the interaction potential to be short ranged:
\begin{align}
 V_{g_1,g_2}(x,y)=0, \text{ if } \sqrt{x^2+y^2} > \xi,
\end{align}
where $\xi$ describes the interaction range.  Hamiltonians with
short-ranged interactions are called local Hamiltonians.

The ground state of the $N$ boson system is an eigenvector of $H$:
\ma{
H(g_1,g_2)\Psi_{g_1,g_2}(\v r_1,...,\v r_N)=E_\text{grnd}(g_1,g_2)
\Psi_{g_1,g_2}(\v r_1,...,\v r_N)
}
with the minimal eigenvalue $E_\text{grnd}(g_1,g_2)$.
The eigenvalues of the Hamiltonian are called energies.

Here we assume that the interaction potential may depend on some parameters
$g_1,g_2$.  As we change \ml{g_1,g_2}, the ground states \ml{\Psi_{g_1,g_2}}
for different \ml{g_1,g_2}'s can some times have similar properties. We say
that those states belong to the same phase.  Some other times, they may have
very different properties.  Then we regard those states to belong to the
different phases.

\subsection{What are quantum phases}

More precisely, quantum phases are defined through quantum phase transitions.
So we first need to define \emph{what are quantum phase transitions?}

\begin{figure}[tb]
\begin{center}
\includegraphics[scale=0.6]{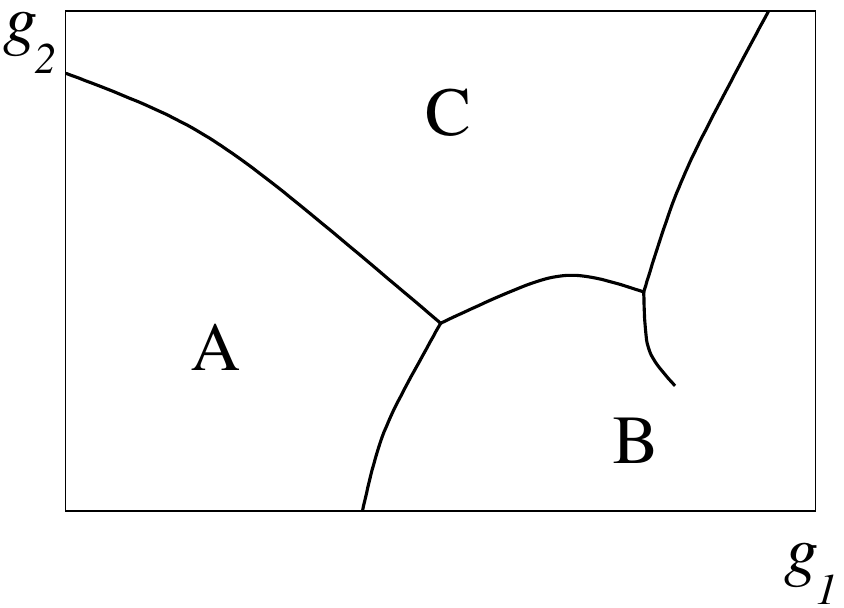}
\end{center}
\caption{
\sffamily
The curves mark the position of singularities in functions
$E_\text{grnd}(g_1,g_2)/N$ and $\<O\>(g_1,g_2)$.  They also represent phase
transitions.  The regions, A, B, and C, separated by phase transitions
correspond to different phases.
}
\label{phaseABC}
\end{figure}

As we change the parameters \ml{g_1,g_2} in the Hamiltonian \ml{H(g_1,g_2)}, if
the average of ground state energy per particle \ml{E_\text{grnd}(g_1,g_2)/N}
has a singularity in \ml{N\to \infty} limit, then the system has a phase
transition.  More generally, if the average of any local operator $O$ on the
ground state:
\begin{align}
\<O\>(g_1,g_2)= 
\int \prod_i \dd x_i \dd y_i \;
\Psi_{g_1,g_2}^* O \Psi_{g_1,g_2},
\end{align}
has a singularity in \ml{N\to \infty} limit as we change $g_1,g_2$, then
the system has a phase transition (see Fig. \ref{phaseABC}).

Using the quantum phase transition, we can define an equivalence relation
between quantum ground states $\Psi_{g_1,g_2}$ in $N\to \infty$ limit: Two
quantum ground states $\Psi_{g_1,g_2}$ and $\Psi_{g_1',g_2'}$ are equivalent if
we can find a path that connect $(g_1,g_2)$ and $(g_1',g_2')$ such that we can
change $\Psi_{g_1,g_2}$ into $\Psi_{g_1',g_2'}$ without encounter a phase
transition.  The quantum phases are nothing but the equivalent classes of such
an equivalence relation.\cite{CGW1038}  In short, the quantum phases are
regions of $(g_1,g_2)$ space which are separated by phase transitions (see Fig.
\ref{phaseABC}).

\subsection{How to classify quantum phases of matter}

One of the most important questions in condensed matter physics is how to
classify the many different quantum phases of matter. One attempt is the theory
of symmetry breaking,\cite{L3726,GL5064,LanL58} which tells us that we should
classify various phases based on the symmetries of the ground state wave
function. Yet with the discovery of the FQH states\cite{TSG8259,L8395} came
also the understanding that there are many distinct and fascinating quantum
phases of matter, called topologically ordered phases,\cite{Wtop,Wrig} whose
characterization has nothing at all to do with symmetry. How should we
systematically classify the different possible topological phases that may
occur in a FQH system?  In this paper, we will try to address this issue.

We know that the FQH states contain topology-dependent degenerate ground
states, which are topologically stable (\ie robust against any  \emph{local} 
perturbations of
the  Hamiltonians).  This allows us to introduce the concept of
topological order in FQH states.\cite{WN9077,W9505} Such
topology-dependent degenerate ground states suggest that the low energy
theories describing the FQH states are topological quantum field
theories\cite{W8951,ZHK8982,FK8967}, which take a form of pure Chern-Simons
theory in 2+1 dimensions.\cite{BW9033,BW9045,R9002,FK9169,WZ9290,FS9333} So one
possibility is that we may try to classify the different FQH phases by
classifying all of the different possible pure Chern-Simons theories.  Although
such a line of thinking leads to a classification of Abelian FQH states in
terms of integer $K$-matrices,\cite{BW9045,R9002,FZ9117,FK9169,WZ9290,FS9333}
it is not a satisfactory approach for non-Abelian FQH
states\cite{MR9162,W9102} because we do not have a good way of knowing which
pure Chern-Simons theories can possibly correspond to a physical system made of
bosons and which cannot. 

Another way to classify FQH states is through the connection between FQH wave
functions and conformal field theory (CFT). It was discovered around 1990 that
correlation functions in certain two-dimensional conformal field theories may
serve as good model wave functions for FQH states.\cite{MR9162,BW9215,WW9455}
Thus perhaps we may classify FQH states by classifying all of the different
CFTs.  However, the relation between CFTs and FQH states is not one-to-one.  If
a CFT produces  a FQH wave function, then any other CFTs that contain the first
CFT can also produce the FQH wave function.\cite{WW9455}

Following the ideas of CFT and in an attempt to obtain a systematic
classification of FQH states without using conformal invariance, it was shown
recently that a wide class of FQH states and their topological
excitations can be classified by their \emph{patterns of zeros}, which describe
the way ideal FQH wave functions go to zero when various clusters of particles
are brought together.\cite{WW0808,WW0809,BW0932,BW1001a} (We would like to
point out that the ``1D charge-density-wave'' characterization of
FQH states\cite{SL0604,BKW0608,BH0802a,S0802,S1002,FS1115} is closely related to
the pattern-of-zeros approach.) This analysis led to the discovery of some new
non-Abelian FQH states whose corresponding CFT has not yet been identified. It
also helped to elucidate the role of CFT in constructing FQH wave functions:
\emph{The CFT encodes the way the wave function goes to zero as various clusters
of bosons are brought together.} The order of these zeros must satisfy certain
conditions and the solutions to these conditions correspond to particular CFTs.
Thus in classifying and characterizing FQH states, one can bypass the CFT
altogether and proceed directly to classifying the different allowed pattern of
zeros and subsequently obtaining the topological properties of the
quasiparticles from the pattern of zeros.\cite{WW0809,BW0932,BW1001a} This
construction can then even be thought of as a classification of the allowed
CFTs that can be used to construct FQH states.\cite{LWW1024} Furthermore, these
considerations give way to a natural notion of which pattern of zeros solutions
are simpler than other ones. In this sense, then, one can see that the
Moore-Read Pfaffian quantum Hall state\cite{MR9162} is the ``simplest''
non-Abelian generalization of the Laughlin state.

\begin{figure}[tb]
\begin{center}
\includegraphics[scale=0.6]{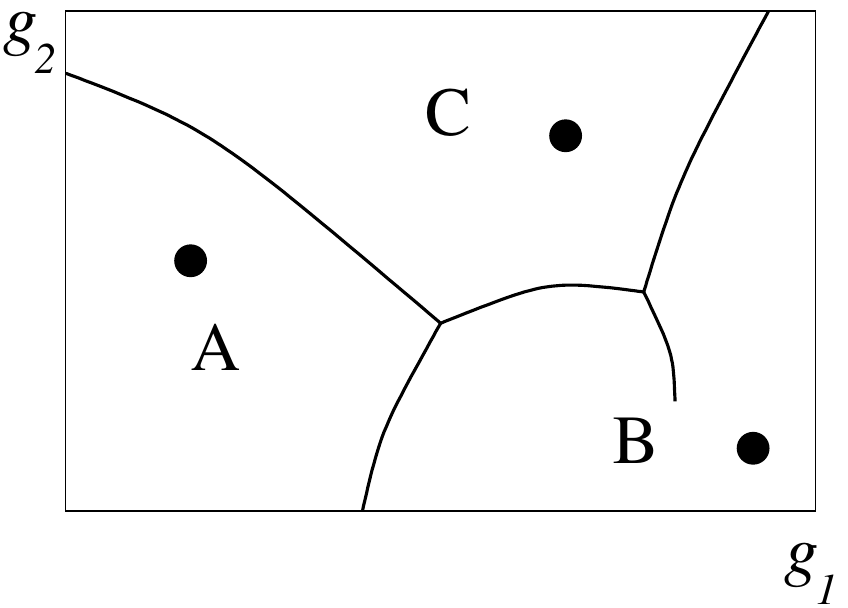}
\end{center}
\caption{
\sffamily
The black dots represent the ideal wave functions that can represent
each quantum phase.
}
\label{phaseABC0}
\end{figure}

We would like to point that in the pattern-of-zeros classification of FQH
states, we do not try to study the phase transition and equivalence
classes.  Instead, we just try to classify some special complex functions of
infinite variables.  We hope those special complex functions can represent each
equivalence class (\ie represent each quantum phase) (see Fig. \ref{phaseABC0}).

\section{Examples of fractional quantum Hall states}

Before trying to classify a type of quantum phases -- FQH phases, let us study
some examples of ideal FQH wave functions to gain some intuitions.

\subsection{The Hamiltonian for FQH systems}

A FQH state of $N$-bosons is described by the following Hamiltonian:
\ma{
H(g_1,g_2)= \sum_{i=1}^N
 (\imth \prt_{z_i} -\imth\frac{1}4 z_i^* )(\imth\prt_{z_i^*} +\imth\frac{1}4 z_i) +\sum_{i<j}
V_{g_1,g_2}(z_i-z_j)
}
where the two dimensional plane is parametrized by \ml{z=x+\imth y}.
When \ml{V_{g_1,g_2}=0}, there are many wave functions
\ma{
\Psi(z_1,\cdots,z_N) = P(z_1,\cdots,z_N)e^{-\frac{1}{4}\sum_{i=1}^N z_iz_i^*},
\ \ \  P =\text{ a symmetric polynomial}
}
that all have the minimal zero eigenvalue (or energy) for any \ml{P}:
\ma{
\Big[\sum_{i=1}^N
(\imth \prt_{z_i} - \imth\frac{1}4 z_i^* )(i\prt_{z_i^*} +\imth\frac{1}4 z_i)\Big]
P(z_1,\cdots,z_N)e^{-\frac{1}{4}\sum_{i=1}^N z_iz_i^*}=0,
}
since
\ma{
e^{\frac{1}{4} zz^*}
(\imth \prt_{z} - \imth \frac{1}4 z^* )(\imth \prt_{z^*} +\imth \frac{1}4 z)
e^{-\frac{1}{4} zz^*}
=(\imth \prt_{z} - \imth \frac{1}2 z^* )\imth \prt_{z^*}
}

For small non-zero  \ml{V_{g_1,g_2}}, there is only one minimal energy wave
function described by a particular polynomial \ml{P} whose form is determined
by \ml{V_{g_1,g_2}}.  In general, it is very hard to calculate this unique
ground state wave function.  In the following, we will show that for some
special interaction potential $V_{g_1,g_2}$, the ground state wave function can
be obtained exactly.

\subsection{Three ideal FQH states:
the exact zero-energy ground states
}

For interaction
\begin{align}
V_{1/2}(z_1,z_2)=\del(z_1-z_2) ,
\end{align}
the wave function 
$P_{1/2}(z_1,\cdots,z_N)e^{-\frac{1}{4}\sum_{i=1}^N z_iz_i^*} $ with
\ma{
P_{1/2}&=\prod_{i<j} (z_i-z_j)^2 
}
is the only zero energy state
with minimal total power of $z_i$'s.
This is because
\begin{align}
\int \prod_i \dd^2 z_i \;
e^{-\frac{1}{4} \sum_i|z_i|^2} 
P_{1/2}^* 
\Big[\sum_{i<j}V_{1/2}(z_i,z_j) \Big]
P_{1/2} 
e^{-\frac{1}{4} \sum_i|z_i|^2} 
=0.
\end{align}
Such a state is called \ml{\nu=1/2} Laughlin state.

For interaction
\begin{align}
V_{1/4}(z_1,z_2) = 
v_0\del(z_1-z_2)+ v_2 \prt^2_{z_1^*} \del(z_1-z_2) \prt^2_{z_1},
\end{align}
the wave function 
$P_{1/4}(z_1,\cdots,z_N)e^{-\frac{1}{4}\sum_{i=1}^N z_iz_i^*} $ with
\ma{
P_{1/4}&=\prod_{i<j} (z_i-z_j)^4 
}
is the only zero energy state
with minimal total power of $z_i$'s,
since
\begin{align}
\int \prod_i \dd^2 z_i \;
e^{-\frac{1}{4} \sum_i|z_i|^2} 
P_{1/4}^* 
\Big[\sum_{i<j}V_{1/4}(z_i,z_j) \Big]
P_{1/4} 
e^{-\frac{1}{4} \sum_i|z_i|^2} 
=0.
\end{align}
Such a state is called \ml{\nu=1/4} Laughlin state. 

Now let us consider interaction\cite{GWW9105,GWW9267}
\begin{align}
V_\text{Pf}(z_1,z_2,z_3) =\cS[
v_0 \del(z_1-z_2) \del(z_2-z_3)
-v_1 \del(z_1-z_2) \prt_{z_3^*} \del(z_2-z_3) \prt_{z_3} ]
\end{align}
where $\cS$ symmetrizes among $z_1,z_2,z_3$
to make $V_\text{Pf}(z_1,z_2,z_3)$ a symmetric function.
Then
the wave function 
$P_\text{Pf}(z_1,\cdots,z_N)e^{-\frac{1}{4}\sum_{i=1}^N z_iz_i^*} $ with
\ma{
P_\text{Pf} =\cA \Big(
\frac{1}{z_1-z_2}
\frac{1}{z_3-z_4}\cdots
\frac{1}{z_{N-1}-z_N}
\Big)
\prod_{i<j} (z_i-z_j)
=
\text{Pf}(\frac{1}{z_i-z_j})
\prod_{i<j} (z_i-z_j)
}
is the only zero energy state
with minimal total power of $z_i$'s,
where $\cA$ anti-symmetrizes among $z_1,...,z_N$.
This is because
\begin{align}
\int \prod_i \dd^2 z_i \;
e^{-\frac{1}{4} \sum_i|z_i|^2} 
P_\text{Pf}^* 
\Big[\sum_{i<j<k}V_\text{Pf}(z_i,z_j,z_k) \Big]
P_\text{Pf} 
e^{-\frac{1}{4} \sum_i|z_i|^2} 
=0.
\end{align}
Such a state is called the Pfaffian state.\cite{MR9162}

\section{The universal properties of FQH phases}

The three many-body wave functions \ml{P_{1/2}e^{-\frac{1}{4} \sum_i|z_i|^2}},
\ml{P_{1/4}e^{-\frac{1}{4} \sum_i|z_i|^2}}, and \ml{P_\text{Pf}e^{-\frac{1}{4}
\sum_i|z_i|^2}} have some amazing  exact properties in $N\to \infty$ limit.  We
believe that those properties do not depend on any local deformations of the
wave functions.\footnote{A local deformation of a many-body wave function
$\Psi$ is generated as
$\Psi\to \Psi'=\e^{\imth \del H}\Psi$ where
$\del H$ is a hermitian operator that can be viewed as an local Hamiltonian.} 
In other words, those properties are shared by all the wave
functions in the same phase.  We call such kind of properties universal
properties.  

The universal properties can be viewed as quantum topological invariants in
mathematics, since they do not change under any perturbations of the local
Hamiltonian.  Thus, from mathematical point of view, the symmetric polynomials
of infinite variables, such as $P_{1/2}$, $P_{1/2}$, and $P_\text{Pf}$, can
have many quantum topological invariants (\ie the universal properties) once we
define their norm to be
\begin{align}
 \<P|P\>=\int \prod_{i=1}^N \dd^2 z_i\; |P(z_1,...,z_N)|^2
\e^{-\frac12 \sum |z_i|^2} .
\end{align}

Since the three wave functions have different universal properties, this
implies that the  three wave functions belong to three different quantum
phases.  In this section, we will discuss some of the  universal properties, by
first listing them in boldface. Then we will give an understanding of them from
physics point of view.  Those conjectured universal properties are exact, but
not rigorously proven to be true.

\subsection{The filling fractions of FQH phases}

\begin{figure}[tb]
\begin{center}
\includegraphics[scale=0.5]{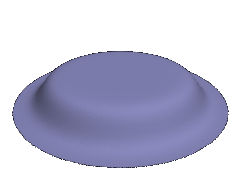}
\end{center}
\caption{
\sffamily
The shape of the density profile $\rho(z)$.
}
\label{disk}
\end{figure}

The density profile of a FQH wave function is is given by
\ma{
\rho(z)=\frac{
\int
d^2z_2 ...  d^2z_N\; |P(z,z_2,...,z_N)|^2e^{-\frac12
\sum|z_i|^2}
}{\int d^2z_1
d^2z_2 ...  d^2z_N\; |P(z_1,z_2,...,z_N)|^2e^{-\frac12
\sum|z_i|^2}
}
}
{\bf 
We believe that
\begin{align}
\nu\equiv 2\pi \rho(0)
\end{align}
is a rational number in $N\to\infty$ limit.}
$\nu$ is called the filling fraction of the corresponding FQH state.
We find that
\begin{align}
P_{1} =\prod (z_i-z_j)\ \ \to \ \ &\nu=1, &
P_{1/2} =\prod (z_i-z_j)^2 \ \ \to \ \ & \nu=1/2,
\nonumber\\
P_{1/4} =\prod (z_i-z_j)^4 \ \ \to \ \ & \nu=1/4, &
P_\text{Pf} =\text{Pf}(\frac{1}{z_i-z_j})\prod (z_i-z_j) \ \ \to \ \ & \nu=1.
\end{align}
Note that $P_{1}$ is anti-symmetric and describe a many-fermion state, while
$P_{1/2}$, $P_{1/4}$, and $P_\text{Pf}$ are symmetric and describe many-boson
states.

We also believe that the density profile $\rho(z)$ has disk shape (see Fig.
\ref{disk}) in large $N$ limit: $\rho(z)$ is almost a constant $\nu/2\pi$
for $|z|< \sqrt{2N/\nu}$ and quickly drop to almost zero for  $|z|>
\sqrt{2N/\nu}$.

\subsubsection{Why $\nu=1$ for state
$\Psi_{1}=\prod_{i<j} (z_i-z_j) \e^{-\sum |z_i|^2/4} $
}
\label{nu1}

\begin{figure}[tb]
\begin{center}
\hfil
\includegraphics[scale=0.4]{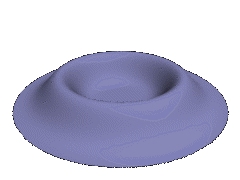}
\hfil
\includegraphics[scale=0.5]{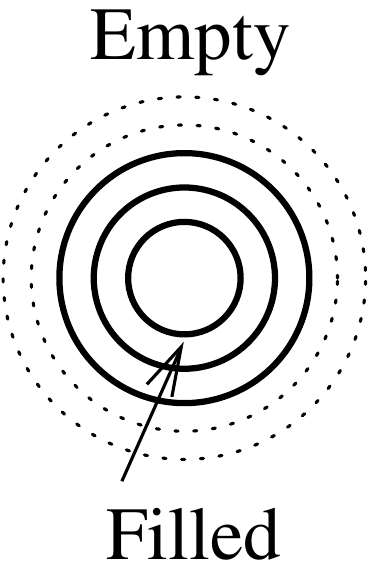}
\hfil
\includegraphics[scale=0.4]{disk}\\
\hfil
(a)
\hfil
~~~~~~~~~
(b)
~~~~~~~~~~
\hfil
(c)
~~~~~~
\end{center}
\caption{
\sffamily
(a) The density profile of the $l^\text{th}$ orbital.
(b) The filling of the orbitals gives rise to a disk-like density profile
in (c).
}
\label{ringdisk}
\end{figure}

We note that the one-particle eigenstates (the orbitals) for one-particle
Hamiltonian \ml{ H_0= -\sum (\prt_z -\frac{1}4 z^* )(\prt_{z^*} +\frac{1}4 z)}
can be labeled by the angular momentum \ml{l}, which is given by \ml{ z^l
e^{-\frac14 |z|^2} }.  The one-particle eigenstate has a ring-like shape with
maximum at \ml{|z|=r_l= \sqrt{2 l}} (see Fig. \ref{ringdisk}a).
The \ml{\nu=1} many-fermion state is obtained by filling the
orbitals (see Fig. \ref{ringdisk}b):
\ma{
\Psi=\prod_{i<j} (z_i-z_j)e^{-\frac14 \sum |z_i|^2}
=\cA[(z_1)^0 (z_2)^1...] e^{-\frac14 \sum |z_i|^2}
}
We see that there are
\ml{l} fermions within radius \ml{r_l}.
So there is one fermion per
\ml{\pi r_l^2/l =2\pi} area, and thus
\ml{\nu=1} (see Fig. \ref{ringdisk}c).

\subsubsection{Why $\nu=1/m$ for the Laughlin state
$\Psi_{1/m}=\prod_{i<j} (z_i-z_j)^m \e^{-\sum |z_i|^2/4} $
}

Let us consider the joint probability distribution of boson positions, which is given 
by the absolute-value-square of the ground state wave function:
\ma{
p(z_1 \cdots z_N)  &\propto
\left| \Psi_{1/m}(z_1 \cdots z_N) \right|^2
\nonumber\\
&=
\e^{-2 m \sum_{i<j} \ln \left| z_i-z_j \right| -
\frac{m}{2} \sum_i \left|  z_i \right|^2 }
=\e^{-\beta V (z_1 \cdots z_N)}
}
 Choosing \ml{T = \frac{1}{\beta} = \frac{m}{2}},
we can view
$\e^{-\beta V (z_1 \cdots z_N)}$ as the 
probability distribution
for $N$ particles
with potential energy $V(z_1 \cdots z_N)$
at temperature $T=\frac{m}{2}$.
The potential has a form
\ma{
V = -m^2 \sum_{i<j} \ln \left| z_i-z_j \right| +
\frac{m}{4} \sum_i \left|  z_i \right|^2
}
which is the potential for a two-dimensional plasma of `charge' \ml{m}
particles.\cite{L8395}  The two-body term $ -m^2 \ln \left| z-z' \right|$
represents the interaction between two particles and the one-body term
$\frac{m}{4}  \left| z \right|^2$ represents the interaction of a particle with
the background ``charge''.
 
For a uniform background ``charge'' distribution with charge density
\ml{\rho_\phi}, a charge \ml{m} particle at \ml{z} feel a force, \ml{F=(\pi
|z|^2 \rho_\phi) (m)/|z|}.  The corresponding background potential energy is
\ml{-\rho_\phi m \frac{\pi}{2} |z|^2}.  We see that to produce the one-body
potential energy $\frac{m}{4}  \left| z \right|^2$ we need to set \ml{\rho_\phi
= - 1/2\pi }.  Since the plasma must be ``charge'' neutral: \ml{m\rho +
\rho_\phi=0}, we find that \ml{\rho= \frac{1}{m} \frac{1}{2\pi} }.  So
$\nu=1/m$.

\subsection{Quasiparticle and Fractional charge in $\nu=1/m$ Laughlin states}
\label{qpsec}

\begin{figure}[tb]
\begin{center}
\includegraphics[scale=0.7]{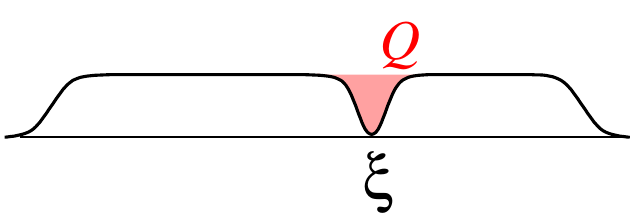} 
\end{center}
\caption{
\sffamily
The density profile of a many-boson wave function with
a quasi-hole excitation at $\xi$.
}
\label{qpart}
\end{figure}

If we remove a boson at position $\xi$ from the Laughlin wave function
$\prod_{i<j} (z_i-z_j)^m \e^{-\sum |z_i|^2/4}$, we create a hole-like 
excitation
described by the wave function $\Psi_\xi^\text{hole}(z_1,...,z_N)$:
\ma{
\Psi_\xi^\text{hole}(z_1,...,z_N)\propto
\prod_i (\xi-z_i)^m \prod_{i<j} (z_i-z_j)^m \e^{-\sum |z_i|^2/4}
}
Despite the hole-like excitation has a charge = 1, the minimal
value for non-zero integers,
it is not the minimally charged excitation.
The minimally charged excitation corresponds to 
a quasi-hole excitation, which is
 described by the wave function
\ma{
\Psi_\xi^\text{quasi-hole}(z_1,...,z_N)\propto
\prod_i (\xi-z_i) \prod_{i<j} (z_i-z_j)^m \e^{-\sum |z_i|^2/4}
}
The density profile
for the quasi-hole wave function $\Psi_\xi^\text{quasi-hole}(z_1,...,z_N)$
is given by 
\begin{align}
\rho_\xi (z)= 
\frac{
\int \prod_{i=2}^N \dd^2 z_i \;
|\Psi_\xi^\text{quasi-hole}(z,z_2,...,z_N)|^2
}{
\int \prod_{i=1}^N \dd^2 z_i \;
|\Psi_\xi^\text{quasi-hole}(z_1,z_2,...,z_N)|^2
}
\end{align}
$\rho_\xi (z)$ has a shape as in Fig. \ref{qpart}.
The quasi-particle charge is defined as
\begin{align}
\label{qcharge}
Q= \int_{D_\xi} \dd^2 z \Big( \frac{\nu}{2\pi}-\rho_\xi (z) \Big) 
\end{align}
in the $N\to \infty$ limit, where $D_\xi$ is a big disk covering $\xi$.  (Note
that, away from the quasi-hole, $\rho_\xi(z)=\frac{\nu}{2\pi}$.) We believe
that {\bf the quasi-hole charge is a rational number  $Q=1/m$}.\cite{L8395}

One way to understand the above result is to note that
\ml{m} quasi-holes correspond to a missing boson:
\ml{\big[\prod_i (\xi-z_i)\big]^m=\prod_i (\xi-z_i)^m}.
So a quasi-hole excitation has a fractional charge $1/m$
although the FQH state is formed by particles of charge 1!

We can also calculate the quasi-hole charge directly.
Note that,
for the Laughlin state
$\Psi_\xi^\text{quasi-hole}(z_1,...,z_N)$
with a quasi-hole
at $\xi$, 
the corresponding joint probability distribution of boson positions
is given by
\ml{p(\{z_i\})\propto |\Psi_\xi^\text{quasi-hole}(\{z_i\}) |=\e^{-\bt V}
} with
\ma{
V =
-m^2 \sum_{i<j} \ln \left| z_i-z_j \right|
-m \sum_{i} \ln \left| z_i-\xi \right|
+ \frac{m}{4} \sum_i \left|  z_i \right|^2
}
Now, the one-body potential term
$-m  \ln \left| z-\xi \right|
+ \frac{m}{4}  \left|  z \right|^2$
is produced by
background charge density: \ml{\rho_\phi=-\frac{1}{2\pi}+
\del(\xi)}.
The 
``charge'' neutral condition \ml{m\rho_\xi (z) +
\rho_\phi(z)\approx 0}
allows us to show that
$\rho_\xi(z)$ has a shape as in Fig. \ref{qpart}
and satisfies \eqn{qcharge} with $Q=1/m$.

\subsection{The concept of quasiparticle type}

We would like to point out that the wave function $
\Psi_\xi^\text{quasi-hole}(z_1,...,z_N)\propto \prod_i (\xi-z_i) \prod_{i<j}
(z_i-z_j)^m \e^{-\sum |z_i|^2/4} $ just describes a particular kind of
quasiparticle excitation.  More general quasiparticle excitations can be
constructed as
\begin{align}
 \Psi_\xi^{\text{quasi-hole-}k}(z_1,...,z_N)\propto \prod_i (\xi-z_i)^k \prod_{i<j}
(z_i-z_j)^m \e^{-\sum |z_i|^2/4}.
\end{align}
which can be viewed as a bound state of $k$ charge-1/m quasi-holes.
So it appears that different types of
quasiparticles are labeled by integer $k$.

Here we would like to
introduce a concept of quasiparticle type: \emph{two quasiparticles
belong to the same type if they only differ by a number of bosons that form the
FQH state.} Since the quasiparticle labeled by $k=m$ correspond to a boson, so
the different types of quasiparticles in the $\nu=1/m$ Laughlin state are
labeled by $k$ mod $m$.  {\bf There are $m$ types  quasiparticles in the
$\nu=1/m$ Laughlin state} (including the trivial type labeled by $k=0$).

There is an amazing relation between the number of quasiparticle type and the
ground state degeneracy of the FQH state on torus: {\bf the number of
quasiparticle type always equal to the ground state degeneracy on torus,
in the $N\to \infty$ limit}.

\subsection{Fractional statistics in Laughlin states }

We note that
the normalized state with a quasi-hole at $\xi$ is described by an \ml{N}-boson 
wave function parameterized by \ml{\xi}:
\ma{
\Psi^\text{quasi-hole}_{\xi}=[N(\xi,\xi^*)]^{-1/2}
\prod_i (\xi-z_i) \prod_{i<j} (z_i-z_j)^2 \e^{-\sum |z_i|^2/4}
}
where $N(\xi,\xi^*)$ is the normalization factor.
The normalized two quasi-hole wave function is given by
\ma{
&\Psi^\text{quasi-hole}_{\xi,\xi'}=[N(\xi,\xi^*,\xi',\xi'^*)]^{-1/2}
\prod_i (\xi-z_i)
\prod_i (\xi'-z_i)
\prod_{i<j}
(z_i-z_j)^2 \e^{-\sum |z_i|^2/4}
}
{\bf 
We conjecture that the above two normalization factors are given by
\begin{align}
\label{Nxi}
N(\xi,\xi^*)=\e^{\frac{1}{2m}|\xi|^2}\times \text{Const.}
\end{align}
and
\begin{align}
\label{Nxixip}
N(\xi,\xi^*,\xi',\xi'^*)
=\e^{
\frac{1}{2m}(|\xi|^2 + |\xi'|^2)
+\frac{1}{m}\ln|\xi-\xi'|^2
}\times
\text{Const.}
\end{align}
in the $N\to \infty$ limit, where $\xi$ and $\xi'$ are hold fixed
in the limit.}

The quasi-holes in the  Laughlin states also have fractional 
statistics.\cite{LM7701,W8257,H8483,ASW8422}  We
can calculate the fractional statistics by calculating the 
Berry phase\cite{B8445} of moving the quasi-holes.  
It turns out that the Berry phase of moving the quasi-holes can be calculated
from the above normalization factors.  Let us first calculate the Berry phase for one
quasi-hole and the normalization factor $N(\xi,\xi^*)$.
The Berry's phase $ \Del \vphi$ induced by moving
$\xi$ is defined as
$ \e^{\imth \Del \vphi}=\<\Psi^\text{quasi-hole}_{\xi}|\Psi^\text{quasi-hole}_{\xi+d\xi}\>$.
It is given by
\ma{
\Del \vphi= a_\xi d\xi +a_{\xi^*} d\xi^* , \ \ \ \
a_{\xi}=-\imth \<\Psi_{\xi}|\frac{\prt}{\prt \xi} |\Psi_{\xi}\>,
\ \
a_{\xi^*}=-\imth \<\Psi_{\xi}|\frac{\prt}{\prt \xi^*} |\Psi_{\xi}\>,
}
where $a_{\xi}$ and $a_{\xi^*}$ are Berry connections.
Since the unnormalized state \ml{\prod_i (\xi-z_i) \prod_{i<j}
(z_i-z_j)^2 \e^{-\sum |z_i|^2/4}} has a special property that it only
depends only on
\ml{\xi} (holomorphic), the Berry connection \ml{(a_\xi,a_{\xi^*})} 
can be calculated from
the normalization \ml{N(\xi,\xi^*)}
of the holomorphic state:
\ma{
a_{\xi}=-\frac{\imth}{2}\frac{\prt}{\prt \xi} \ln [N(\xi,\xi^*)],\ \ \ \
a_{\xi^*}=\frac{\imth}{2}\frac{\prt}{\prt \xi^*} \ln [N(\xi,\xi^*)] .
}

Now let us calculate $N(\xi,\xi^*)$.  Let us guess that
\ml{N(\xi,\xi^*)} is given by \eqn{Nxi}.  To show the guess to be right,
we need to show that
the norm of $|\Psi^\text{quasi-hole}_\xi\>$ does not depend on
$\xi$.  We note that \ml{ |\Psi^\text{quasi-hole}_\xi|^2 = \e^{-\bt V_\xi} } with
\ma{
V_\xi(z_1,...,z_N) =
-m^2 \sum_{i<j} \ln \left| z_i-z_j \right|
-m \sum_{i} \ln \left| z_i-\xi \right|
+ \frac{1}{4} \left|  \xi \right|^2
+ \frac{m}{4} \sum_i \left|  z_i \right|^2.
}
Here $V_\xi$
can be viewed as the total energy of a plasma
of \ml{N} `charge'-\ml{m} particles at \ml{z_i}
and one `charge'-\ml{1} particle hold fixed at \ml{\xi}.
Both particles interact with the same background charge.
Note that
the norm $\<\Psi^\text{quasi-hole}_\xi|\Psi^\text{quasi-hole}_\xi\>$
is given by
\begin{align}
\<\Psi^\text{quasi-hole}_\xi|\Psi^\text{quasi-hole}_\xi\>= \int \prod \dd^2 z_i\;  \e^{-\bt V_\xi}
\end{align} 
 Due to the screening of the plasma,
we argue that $ \int \prod \dd^2 z_i\;  \e^{-\bt V_\xi} $
does not depend on \ml{\xi} in $N\to \infty$ limit,  which implies that
\ml{ \<\Psi^\text{quasi-hole}_\xi|\Psi^\text{quasi-hole}_\xi\> } does not depend on \ml{\xi}.
Thus 
\ml{N(\xi,\xi^*)} is indeed given by \eqn{Nxi}.

This allows us to find
\ma{
a_\xi=-\imth \frac{1}{4m} \xi^*,\
a_{\xi^*}= \imth \frac{1}{4m} \xi
}
Using such a Berry connection, let us calculate
the Berry's phase for moving $\xi$ around a circle $C$
of radius $r$ center at $z=0$:
\ma{
\label{vphi}
\Del \vphi=
& \oint_C (a_\xi d\xi+ a_{\xi^*} d\xi^*)
=2\pi \frac{ r^2}{4 m}\times 2
=2\pi \frac{ \text{Area enclosed by }C}{2\pi m}
\nonumber\\
&=2\pi\times \text{number of enclosed bosons by }C.
}
We see that the Berry connection describes a uniform `magnetic' field.  The
above result can also be understood directly from the wave function \ml{\prod_i
(\xi-z_i) \prod_{i<j} (z_i-z_j)^2 \e^{-\sum |z_i|^2/4}}.

%

Similarly, we can calculate the
Berry connection for two quasi-holes. 
Let us guess that
$N(\xi,\xi^*,\xi',\xi'^*)$
is given by \eqn{Nxixip}.
For such a normalization factor, we find that
$|\Psi^\text{quasi-hole}_{\xi,\xi'}|^2=\e^{-\bt V_{\xi,\xi'}}$ with
\begin{align}
&V_{\xi,\xi'}(z_1,...,z_N) =
-m \sum_{i}
[\ln \left| z_i-\xi \right|
+\ln \left| z_i-\xi' \right|
]
+ \frac{1}{4} [
\left|  \xi \right|^2
+\left|  \xi' \right|^2
]
- \ln \left| \xi-\xi' \right|
\nonumber\\
&\ \ \ \ \ \ \ \ \ \ \
-m^2 \sum_{i<j} \ln \left| z_i-z_j \right|
+ \frac{m}{4} \sum_i \left|  z_i \right|^2
\end{align}
Such a $V_{\xi,\xi'}$ can be viewed as the total energy of a plasma of \ml{N}
`charge'-\ml{m} particles at \ml{z_i} and two `charge'-\ml{1} particles at
\ml{\xi} and \ml{\xi'}.  Due to the screening,  $ \int \prod \dd^2 z_i\;  \e^{-\bt V_{\xi,\xi'}} $ does not
depend on \ml{\xi} and \ml{\xi'} in $N\to \infty$ limit, which implies that \ml{
\<\Psi^\text{quasi-hole}_{\xi,\xi'}|\Psi^\text{quasi-hole}_{\xi,\xi'}\> } does not depend on \ml{\xi} and
\ml{\xi'}.
So our guess is correct.
Using the
 normalization factor \eq{Nxixip},
we find the Berry connection to be
\begin{align}
a_\xi=-\imth \frac{1}{4m} \xi^*
+\frac{\imth}{2m}\frac{1}{\xi-\xi'}
,\ \ \ \ \ \
a_{\xi^*}= \imth \frac{1}{4m} \xi
-\frac{\imth}{2m}\frac{1}{\xi^*-\xi'^*}
\end{align}

Using such a Berry connection, we can calculate the fractional statistics of 
the quasi-holes in the $\nu=1/m$ Laughlin state.
 Moving a quasi-hole around another, we find the
Berry phase to be \ml{\Del \vphi=
\frac{
\text{enclosed area}}{m}-\frac{2\pi}{m}
} (see \eqn{vphi} for comparison).
If we only look at the sub-leading term \ml{-2\pi/m}, we find that exchanging
two quasi-holes 
give rise to phase \ml{\th=-\pi/m}, since
exchanging
two quasi-holes
correspond to moving a quasi-hole half way around another 
and we get the half of \ml{-2\pi/m}.
We find that
{\bf quasi-holes in the $\nu=1/m$ Laughlin state have a
fractional statistics
described by the  phase factor \ml{\e^{-\imth\pi/m}}}.\cite{H8483,ASW8422}

The term \ml{\frac{
\text{enclosed area}}{m}} implies that the quasi-holes sees
a uniform magnetic field.
So the  quasi-holes in the $\nu=1/m$ Laughlin state are
anyons in magnetic field.

\subsection{Quasi-holes in the $\nu=1$ Pfaffian state}

\subsubsection{Charge-1 and charge-1/2 quasi-holes}

Ground state wave function for the $\nu=1$ Pfaffian state
is given by
\ma{
\Psi_\text{Pf} &=\cA \Big(
\frac{1}{z_1-z_2}
\frac{1}{z_3-z_4}\cdots
\frac{1}{z_{N-1}-z_N}
\Big) \Psi_1
= \text{Pf} \Big(
\frac{1}{z_i-z_j}
\Big) \Psi_1
}
where $\Psi_1$ is given by $\prod_{i<j} (z_i-z_j) \e^{-\frac14 \sum_i |z_i|^2}$.
 A simple quasi-hole state is given by
\ma{
& 
\Psi_\xi^\text{charge-1} =
\prod (\xi-z_i) \Psi_\text{Pf}
\nonumber\\
&=
\cA \Big(
\frac{(\xi-z_1)(\xi-z_2)}{z_1-z_2}
\frac{(\xi-z_3)(\xi-z_4)}{z_3-z_4}\cdots
\Big)\Psi_1
=
\text{Pf} \Big(
\frac{(\xi-z_i)(\xi-z_j)}{z_i-z_j}
\Big)\Psi_1
}
which is created by multiplying the factor
$\prod (\xi-z_i)$ to the ground state wave function.
{\bf Such a quasi-hole has a charge 1.}
The above quasi-hole can be splitted into two
fractionalized quasi-holes.
A state with two fractionalized quasi-holes
at $\xi$ and $\xi'$
is given by
\ma{
\Psi_{\xi,\xi'}^\text{charge-1/2} &=
\cA \Big(
\frac{(\xi-z_1)(\xi'-z_2)+(1 \leftrightarrow 2)}{z_1-z_2}\
\frac{(\xi-z_3)(\xi'-z_4)+(3 \leftrightarrow 4)}{z_3-z_4}\cdots
\Big) \Psi_1
\nonumber\\
&=
\text{Pf}\Big(
\frac{
(\xi-z_i)(\xi'-z_j)
+(\xi-z_j)(\xi'-z_i)
}{z_i-z_j}
\Big) \Psi_1
}
{\bf Such a fractionalized quasi-hole has a charge 1/2.}
We note that
combining two charge-1/2 quasi-holes give us one charge-1
quasi-hole:
\begin{align}
 \Psi_{\xi,\xi}^\text{charge-1/2} \propto
 \Psi_{\xi}^\text{charge-1}.
\end{align}

\subsubsection{How many states with four charge-1/2 quasi-holes?}

 One of the state with four charge-1/2 quasi-holes
at $\xi_1$, $\xi_2$, $\xi_3$, and $\xi_4$ is given by
\ma{
P_{(12)(34)} &=
\text{Pf}\Big(
\frac{
(\xi_1-z_i)
(\xi_2-z_i)
(\xi_3-z_j)
(\xi_4-z_j)
+(i \leftrightarrow j)
}{z_i-z_j}
\Big) \Psi_1
\nonumber\\
&=
\text{Pf}\Big(
\frac{
[12,34]_{z_iz_j}
}{z_i-z_j}
\Big) \Psi_1
}
The other two are \ml{P_{(13)(14)}}, \ml{P_{(14)(23)}}.
But only two of them are linearly independent.\cite{NW9629}
Using the relation
\ma{
[12,34]_{z_iz_j}-
[13,24]_{z_iz_j}=
(z_i-z_j)^2(\xi_1-\xi_4)(\xi_2-\xi_3)
=z_{ij}^2\xi_{14}\xi_{23}
}
we find (with 
\ml{z_{12}=z_1-z_2}, 
\ml{\xi_{12}=\xi_1-\xi_2}, 
etc)
\ma{
P_{(13)(24)} &=
\cA\Big(
\frac{ [12,34]_{z_1z_2}-z_{12}^2\xi_{14}\xi_{23} }{z_{12} }
\
\frac{ [12,34]_{z_3z_4}-z_{34}^2\xi_{14}\xi_{23} }{z_{34} }
\cdots
\Big)\Psi_1
\nonumber\\
&=
P_{(12)(34)}-N_{pair}\cA\Big(
z_{12}\xi_{14}\xi_{23}
\frac{ [12,34]_{z_3z_4}}{z_{34} }
\cdots
\Big)\Psi_1
}
So
\ma{
P_{(12)(34)}-P_{(13)(24)}=
N_{pair}\xi_{14}\xi_{23} \cA\Big( z_{12}
\frac{ [12,34]_{z_3z_4}}{z_{34} } \cdots \Big) \Psi_1
}
Similarly
\ma{
P_{(12)(34)}-P_{(14)(23)}=
N_{pair}\xi_{13}\xi_{24} \cA\Big( z_{12}
\frac{ [12,34]_{z_3z_4}}{z_{34} } \cdots \Big) \Psi_1
}
Thus
\ma
{
\frac{
P_{(12)(34)}-P_{(13)(24)}
}{\xi_{14}\xi_{23}}
=
\frac{
P_{(12)(34)}-P_{(14)(23)}
}{
\xi_{13}\xi_{24}
}
}
We find that {\bf there are two states for four charge-1/2 quasi-holes, even if
we fixed their positions. The two states are topologically degenerate (have the
same energy in $N\to \infty$ limit).}\cite{NW9629}  The appearance of the
topological degeneracy even with fixed quasi-hole positions is a defining
property of the non-Abelian statistics.  In the presence of the topological
degeneracy, as we exchange quasi-holes, we will generate non-Abelian Berry
phases which also describe non-Abelian statistics.

More generally we find that there are \ml{D_n=\frac 12 (\sqrt 2)^n}
topologically degenerate states for \ml{n} charge-1/2 quasi-holes, even if we
fixed their positions.\cite{NW9629}  We see that there are \ml{ \sqrt 2} states
per charge-1/2 quasi-hole!  The $\sqrt 2$ is called the quantum dimension for
the charge-1/2 quasi-hole.  We see that the charge-1/2 quasi-hole has a
non-Abelian statistics, since for Abelian anyons, the quantum dimension is
always 1.

\subsection{Edge excitations and conformal field theory}

Under the $z\to \e^{\imth \th} z$ transformation, the $N$-particle $\nu=1/2$
Laughlin wave function $\Psi_{1/2}=P_{1/2}(z_1,...,z_N) e^{-\sum |z_i|^2/4} =
\prod_{1\leq i< j\leq N} (z_i-z_j)^2 e^{-\sum |z_i|^2/4}$ transforms as
$\Psi_{1/2} \to \e^{\imth S_N \th} \Psi_{1/2}$, with $S_N=N(N-1)$.  We call
$S_N$ the angular momentum of the Laughlin wave function (which is also the
total power of $z_i$'s of the polynomial $P_{1/2}(z_1,...,z_N)$).  For
interaction $V_{1/2}=\sum \del(z_i-z_j)$, the $\nu=1/2$ Laughlin wave function
is the only zero energy state with angular momentum $N(N-1)$ since $\Psi_{1/2}
(z_1,...,z_N)$ vanishes as $z_i\to z_j$.  There are no zero energy states with
angular momentum less then $S_N$. 
In fact, we believe that,
{\bf for wave functions $\Psi$ with angular momentum less then $S_N$,
\begin{align}
 \<V_{1/2}\> =\frac{
\int \prod \dd^2 z_i\; V_{1/2} |\Psi(z_1,...,z_N)|^2 
}{
\int \prod \dd^2 z_i\; |\Psi(z_1,...,z_N)|^2 
} \geq \Del 
\end{align}
for a positive $\Del$ and any $N$.}
The maximal $\Del$ is called the energy gap for the interaction
$V_{1/2}$.

On the other hand, 
there are many zero energy states ($\<V_{1/2}\>=0$) with angular momentum
bigger than $S_N$.  We call those zero energy states edge states, and denote
them as $\Psi_\text{edge}$.  We can introduce a sequence of integers
$D^\text{edge}_L$ to denote the number of zero energy states with angular
momentum $S_N+L$.  We will call $D^\text{edge}_L$ the edge spectrum.

To obtain the edge spectrum for the \ml{\nu=1/2} Laughlin state with
interaction $V_{1/2}$, we note that the zero-energy edge states can be obtained
by multiplying the Laughlin wave function by a symmetric polynomial which does
not reduce the order of zeros: 
\ma{
\Psi_\text{edge}=P_\text{sym}(\{z_i\}) \Psi_{1/2} 
}
Since the number of the symmetric polynomial
with the total power of $z_i$'s equal to $L$ is given by
the partition number $p_L$, we find $D^\text{edge}_L=p_L$.
Such an argument applies to any Laughlin states. So
we believe that {\bf for $\nu=1/m$ Laughlin
the edge spectrum is given by the partition numbers: 
$D^\text{edge}_L=p_L$}:\cite{W9211}
\begin{align}
\begin{tabular}{|r||c|c|c|c|c|c|c|}
\hline
\ml{L} &       ~0~ & 1 &2 & ~3~ & ~4~ & ~5~ & ~6~ \\
\hline
\ml{D^\text{edge}_L} &  \ml{1}  & \ml{1}     &\ml{2}     & \ml{3}    &\ml{5} 
& 7 & 11 \\
\hline
\ml{P_\text{sym}} &\ml{1}&\ml{\sum z_i}&\ml{(\sum z_i)^2}&\ml{...}&\ml{...}
& ... & ...\\
             &      &             &\ml{\sum z_i^2}  &\ml{...}&\ml{...}
& ... & ...\\
\hline
\end{tabular}
\end{align}
In large $L$ limit,
\ml{
D^\text{edge}_L
\approx \frac{1}{4\sqrt{3} L} \e^{\pi \sqrt{ \frac{2L}{3} } }
\approx \e^{\pi \sqrt{ \frac{2L}{3} } }
}.

For the \ml{\nu=1} Pfaffian state with
the ideal Hamiltonian \ml{\cS[ v_0 \del(z_1-z_2)
\del(z_2-z_3) -v_1 \del(z_1-z_2) \prt_{z_3^*} \del(z_2-z_3)
\prt_{z_3} ] }, \ml{ \Psi_\text{Pf}=\cA\big( \frac{1}{z_1-z_2}
\frac{1}{z_3-z_4} \cdots \big)\prod_{i<j} (z_i-z_j)}, is the
zero-energy state with the minimal total angular momentum $S_N$.
 Other zero-energy states with higher
angular momenta
are given by
\ma{ 
\Psi_\text{edge}= \cA \left(P_\text{any}(\{z_i\})
\frac{1}{z_1-z_2} \frac{1}{z_3-z_4}...\right)
\Psi_1, 
}
where \ml{P_\text{any}} is any polynomial.
Now the counting is much more difficult, since linearly independent 
\ml{P_\text{any}}'s may generate linearly dependent wave functions.
We find, for large even total boson number $N$,
the edge spectrum is given by\cite{W9355}
\begin{align}
\begin{tabular}{|r||c|c|c|c|c|c|c|}
\hline
\ml{L} &       ~0~ & 1 &2 & ~3~ & ~4~ & ~5~ & ~6~ \\
\hline
\ml{D^\text{edge}_L} &  \ml{1}  & \ml{1}     &\ml{3}     & \ml{5}    &\ml{10} 
& 16 & 28 \\
\hline
\end{tabular}
\end{align}
We believe that, {\bf for the \ml{\nu=1} Pfaffian state, the edge spectrum in
large $L$ limit is given by \ml{ D^\text{edge}_L \approx \e^{\pi \sqrt{ 
\frac{2L}{3} } \sqrt c } } with $c=3/2$, if $N\to \infty$ and $L\ll N$.}

It turns out that the edge spectrum for $\nu=1/m$ Laughlin state can be
produced by a central charge $c=1$ CFT and the edge spectrum for \ml{\nu=1}
Pfaffian state can be produced by a central charge $c=3/2$
CFT.\cite{W9211,W9355}  This allows us to connect the edge excitations of a FQH
state to a CFT. 

Using the quasi-hole wave function $\Psi^\text{quasi-hole}_\xi(z_1,...,z_N)$
that describes a quasi-hole at $\xi$, we can even calculate the correlation
function of the quasi-hole operator.  We know that the circular quantum Hall
droplet has a radius $R=\sqrt{2N/\nu}$.  The quasi-hole correlation function on
the edge of the droplet is given by
\begin{align}
 G^\text{quasi-hole}(\th'-\th)
\propto
\int \prod \dd^2 z_i\; 
[\Psi^\text{quasi-hole}_{\xi'}(z_1,...,z_N)]^*
\Psi^\text{quasi-hole}_\xi(z_1,...,z_N)
\Big|_{\xi=R\e^{\imth \th}; \xi'=R\e^{\imth \th'} } .
\end{align}
We find that {\bf
$G^\text{quasi-hole}(\th-\th')$ has a form
\begin{align}
 G^\text{quasi-hole}(\th)
\propto \e^{\imth Q \nu^{-1} N \th}
\Big( \frac{1}{1-\e^{-\imth \th} } \Big)^{2h}
\end{align}
where $Q$ is the quasi-hole charge and $h$ is a rational number.} We will call
$h$ the scaling dimension of the quasi-hole.  {\bf For the $\nu=1/m$ Laughlin
state, we find that $h=\frac{1}{2m}$ for the charge $Q=1/m$ quasi-hole.  For
the $\nu=1$ Pfaffian state, we find that $h=\frac{1}{2}$ for the charge-1
quasi-hole, and $h=\frac{3}{16}$ for the charge-1/2 quasi-hole, all in $N\to
\infty$ limit.}\cite{W9038,W9355}

\section{Pattern-of-zeros approach to FQH states and symmetric polynomials}

Using $P_{1/2}$, $P_{1/4}$, and $P_\text{Pf}$ as examples, we have seen that
symmetric polynomials with infinite variables can have some amazing universal
properties, once we defined the norm of the infinite-variable polynomials to be
$ \int \prod \dd^2 z_i\ |P|^2\e^{-\frac12 \sum |z_i|^2}$.  This suggests that it
may be possible to come up with a definition of `` infinite-variable
symmetric polynomials''.  Such properly defined infinite-variable symmetric polynomials should have
those  amazing universal properties.  The  proper definition also allow us to
classify infinite-variable symmetric polynomials, which will lead to a classification of
FQH phases.

In this section, we will first discuss an attempt to define infinite-variable
symmetric polynomials through pattern of zeros. Then, we will try to provide a
classification of patterns of zeros.  After that, we will use the patterns of
zeros to calculate the universal properties of the corresponding
infinite-variable symmetric polynomials.

\subsection{What is infinite-variable symmetric polynomial}

The main difficulty to define symmetric polynomial with infinite variables is
that the number of the variables is not fixed.  To overcome this difficulty, we
will characterize the symmetric polynomials through their ``local properties''
that do not depend on the number of the variables. One such ``local property''
is pattern of zeros.

\subsubsection{What is pattern of zeros?}

We have seen that the
different short-range interactions \ml{V(z_i-z_j)} in
Hamiltonian
\ma{
H= \sum_{i=1}^N
- (\prt_z -\frac{B}4 z^* )(\prt_{z^*} +\frac{B}4 z) +\sum_{i<j}
V_(z_i-z_j)
}
leads to different FQH states $P(z_1,...,z_N)\e^{-\frac14 \sum_{i=1}^N
|z_i|^2}$, which in turn leads to different symmetric polynomials
$P(z_1,...,z_N)$.

One of the resulting polynomial 
\ml{
P_{1/2}=\prod_{i<j} (z_i-z_j)^2 
}
has a property that as
 \ml{z_1\approx z_2}, it
has a second-order zero $P_{1/2} \propto (z_1-z_2)^2$.
Another resulting polynomial 
\ml{
P_{1/4}=\prod_{i<j} (z_i-z_j)^4
}
has a property that as
 \ml{z_1\approx z_2}, it
has a fourth-order zero $P_{1/4} \propto (z_1-z_2)^4$.
The third resulting polynomial 
\ma{
P_\text{Pf} &=\cA \Big(
\frac{1}{z_1-z_2}
\frac{1}{z_3-z_4}\cdots
\frac{1}{z_{N-1}-z_N}
\Big)
\prod_{i<j} (z_i-z_j)
}
has a property that as
 \ml{z_1\approx z_2}, $P_\text{Pf}$ has no zero, while as
\ml{z_1\approx z_2\approx z_3}, $P_\text{Pf}$ has a second-order zero.
We see that different polynomials can be
characterized by different pattern of zeros.

The above examples suggest
the following general definition of pattern
of zeros for a symmetric polynomial $P(\{z_i\})$.
Let \ml{z_i=\la \eta_i+z^{(a)}}, \ml{i=1,2,\cdots,a}.
In the small $\la$ limit, we have
\ma{
P(\{z_i\})=\la^{S_a} P(\eta_1,...,\eta_a;z^{(a)},z_{a+1},z_{a+2},\cdots) + O(\la^{S_a+1})
}
 The sequence of integers \ml{\{S_a\}} characterizes the symmetric polynomial
\ml{P(\{z_i\})} and is called the \emph{pattern of zeros} of $P$.  We note that
\ml{S_N} happen to be the total power of \ml{z_i} (or the total angular
momentum) of $P$ if the polynomial has \ml{N} variables.

\subsubsection{The unique fusion condition}

If the above induced $P(\{\eta_i\};z^{(a)},z_{a+1},z_{a+2},\cdots)$,
does not depend on
the ``shape'' \ml{\{\eta_i\}} 
\ma{
P(\{\eta_i\};z^{(a)},z_{a+1},z_{a+2},\cdots)\propto
P(z^{(a)},z_{a+1},z_{a+2},\cdots),
}
we then say that the symmetric polynomial $P(\{z_i\})$ satisfy the \emph{unique
fusion condition}.

\subsubsection{Different encodings of pattern of zeros $S_a$}

There are many different ways to encode the sequence of integers $S_a$.
For example, we may use
\begin{align}
l_a=S_a-S_{a-1}, \ \ \ \ a=1,2,3,...
\end{align}
to encode $S_a$, $a=1,2,3,...$:
\begin{align}
 S_a=\sum_{i=1}^a l_i.
\end{align}
Here we have assumed that $S_0=0$.
It turns out that $l_i\geq 0$ and
$l_i\leq l_{i+1}$.

We may also use $n_l$, $l=0,1,2,...$ to encode $S_a$.
Here $n_l$ is the number of times
that the value $l$ appears in the sequence $l_i$:
\begin{align}
 n_l=\sum_{i=1}^\infty \del_{l,l_i}.
\end{align}

Let us list the pattern of zeros for some simple
polynomials.
For the \ml{\nu=1} integer quantum Hall state 
$P_1=\prod_{i<j}(z_i-z_j)$, 
the pattern of zeros is
is given by
\ma{
S_1,S_2,\cdots:&\ 0,1,3,6,10,15,\cdots
\nonumber\\
l_1,l_2,\cdots:&\ 0,1,2,3,4,5,\cdots
\nonumber\\
n_0n_1n_2\cdots:&\ 11111111\cdots
}
We see that we can view $l$ in $n_l$ as the label for the orbital
$z^l\e^{-\frac14 |z|^2}$, and $n_l$ as the occupation number on the
$l^\text{th}$ orbital (see section \ref{nu1} and Fig. \ref{ringdisk}b).

The pattern of zeros of \ml{\nu=1/2} Laughlin state $P_{1/2}$
is described by
\ma{
S_1,S_2,\cdots:&\ 0,2,6,12,20,30,\cdots
\nonumber\\
l_1,l_2,\cdots:&\ 0,2,4,6,8,10,\cdots
\nonumber\\
n_0n_1n_2\cdots:&\ 1010101010101010\cdots
}
We see that $n_l$ has a periodic structure. Each unit cell
(each cluster) has \ml{1} particle and \ml{2} orbitals

The pattern of zeros of \ml{\nu=1/4} Laughlin state $P_{1/4}$
is described by
\ma{
S_1,S_2,\cdots: &\ 0, 4, 12, 24, 40, 60, 84, \cdots
\nonumber\\
l_1,l_2,\cdots:&\ 0,4,8,12,16,20,\cdots
\nonumber\\
n_0n_1n_2\cdots:&\ 100010001000100010001\cdots
}
Again,  $n_l$ has a periodic structure. Each unit cell
(each cluster) has \ml{1} particle and \ml{4} orbitals

For the \ml{\nu=1} Pfaffian state
\ml{P_\text{Pf}=
\cA \Big(
\frac{1}{z_1-z_2}
\frac{1}{z_3-z_4}\cdots
\Big)
\prod_{i<j} (z_i-z_j)
},
the pattern of zeros is given by
\ma{
S_1,S_2,\cdots: &\ 0,0, 2, 4, 8, 12, 18, 24, \cdots
\nonumber\\
l_1,l_2,\cdots:&\ 0,0,2,2,4,4,6,6,\cdots
\nonumber\\
n_0n_1n_2\cdots:&\ 2020202020202020202\cdots
}
Now a cluster (unit cell) has \ml{2} particles and \ml{2} orbitals.

\subsubsection{The cluster condition}

Motivated by the above examples, here we would like introduce a cluster
condition for symmetric polynomials: \emph{an symmetric polynomial
satisfies a cluster condition if $n_l$ is periodic.}
Let each unit cell contains $n$ particles and $m$ orbitals.
In this case, $S_a$ has a form
\ma{
\label{clusterC}
S_{a+kn} &= S_a+kS_n +\frac{k(k-1)nm}{2} +kma 
}
Since \ml{S_1=0}, we see that we can use a finite sequence
\ml{(\frac{m}{n};S_2,\cdots, S_n)} to describe
the pattern of zeros for symmetric polynomial satisfying the
cluster condition.

We note that the filling fraction $\nu$ is given by the average number of
particles per orbital. Thus  $\nu=n/m$.  We also call the cluster condition
with $n$ particles per unit cell an $n$-cluster condition.

\subsubsection{A definition of infinite-variable symmetric polynomial}

Now, we are ready to define \emph{the infinite-variable symmetric polynomial as
a symmetric polynomial that  satisfies the unique fusion condition and the
cluster condition}.  The cluster condition makes the $N\to \infty$ limit
possible.

From the above discussions, we see that an infinite-variable symmetric
polynomial can be described by a finite amount of data $(\frac{m}{n};S_2,\cdots,
S_n)$.  The \ml{\nu=1/2} Laughlin state, $P_{1/2}$, satisfies the unique fusion
condition and cluster condition.  So $P_{1/2}$ is an infinite-variable
symmetric polynomial described by a pattern of zero:
\ml{(\frac{m}{n};S_2,\cdots, S_n)=(\frac{2}{1};)}.

\subsection{A classification of infinite-variable symmetric polynomials}

We have seen that each infinite-variable symmetric polynomial \ml{P(\{z_i\})}
has a sequence of integers \ml{\{S_a\}} -- a pattern of zeros.  But each
sequence of integers  \ml{\{S_a\}} may not correspond to an infinite-variable
symmetric polynomial \ml{P(\{z_i\})}. In this subsection, we will try to find
all the conditions that a sequence \ml{\{S_a\}} must satisfy, such that
\ml{\{S_a\}} describes a infinite-variable symmetric polynomial.  This may
lead to a classification of infinite-variable symmetric polynomials (or FQH
states) through pattern of zeros.

\subsubsection{Derived polynomials}

To find the conditions on $\{S_a\}$, it is very helpful to introduce
the derived polynomials.
 Let \ml{z_1,...,z_a \to z^{(a)}} in
an infinite-variable symmetric polynomial \ml{P(\{z_i\})} and
use the unique fusion condition:
\ma{
P(\{z_i\})\to \la^{S_a} P_\text{derived}(z^{(a)},z_{a+1},z_{a+2},\cdots) + O(\la^{S_a+1}) ,
}
we obtain a derived polynomial $P_\text{derived}(z^{(a)},z_{a+1},z_{a+2},\cdots)$ from the original polynomial $P$.
Repeating the process on other variables, we get a more general  derived
polynomial \ml{P_\text{derived}(z^{(a)}, z^{(b)}, z^{(c)}, \cdots) }, where
$z^{(a)}$, $z^{(b)}$, \etc are fusions of $a$ variables, $b$ variables, \etc.

The zeros in derived polynomials are described by \ml{D_{a,b}}:
\ma{
P_\text{derived}(z^{(a)}, z^{(b)}, z^{(c)}, \cdots)
\sim (z^{(a)}- z^{(b)})^{D_{a,b}} P_\text{derived}'(z^{(a+b)}...) +\cdots
}
where $z^{(a+b)}=(z^{(a)}+ z^{(b)})/2$.
$D_{a,b}=D_{b,a}$ also characterize the pattern of zeros.
In effect,
\ml{D_{a,b}} and \ml{S_a} encode the same information:
\ma{
D_{a,b}=S_{a+b}-S_a-S_b, 
\ \ \ \
S_a=\sum_{b=1}^{a-1} D_{b,1}
.
}


\subsubsection{The concave conditions on pattern of zeros}

Since $D_{a,b} \geq 0$, we obtain the first concave condition:
\ma{
\label{Del2}
\Del_2(a,b)&\equiv S_{a+b}-S_a-S_b \geq 0 .
}
Such a condition comes from the fusion of two clusters.
We also have a second  concave condition:
\ma{
\label{Del3}
\Del_3(a,b,c) &\equiv S_{a+b+c} -S_{a+b}-S_{b+c}-S_{a+c} +S_a+S_b+S_c \geq 0
}
from the fusion of three clusters.

To derive the second  concave condition, let us fix all variables $z^{(b)},
z^{(c)}, ...$ except $z^{(a)}$ in the derived polynomial
$P_\text{derived}(z^{(a)}, z^{(b)}, z^{(c)}, \cdots)$.  Then the derived
polynomial $P_\text{derived}(z^{(a)}, z^{(b)}, z^{(c)}, \cdots)$ can be viewed
as a complex function $f(z^{(a)})$, which has isolated on-particle zeros at
$z^{(b)}$, $z^{(c)}$, ..., and possibly some other off-particle zeros.

\begin{figure}[tb]
\centerline{
\includegraphics[scale=0.6]{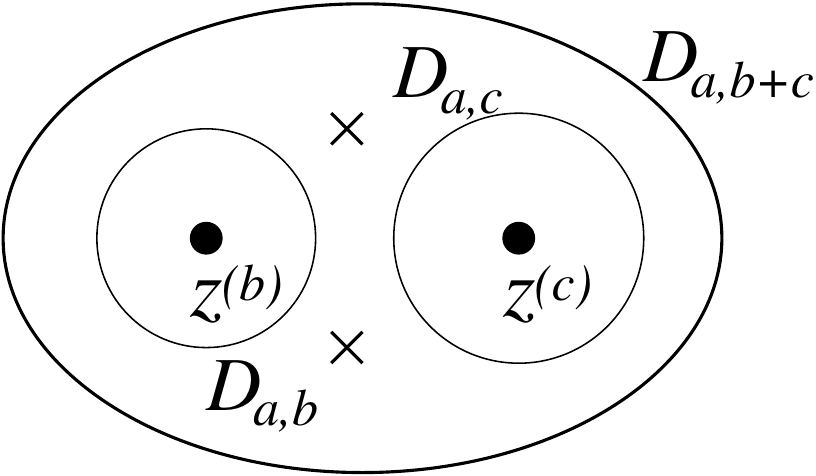}
}
\caption{
\sffamily
$W_{a,bc}$ obtained by moving $z^{(a)}$ along a large loop around
$z^{(b)}$ and
$z^{(c)}$ counts the total numbers of zeros of $f(z^{(a)})$ in the loop.
The crosses mark the off-particle zeros of $f(z^{(a)})$ not at $z^{(b)}$ and
$z^{(c)}$.
}
\label{DDD}
\end{figure}

Let us move $z^{(a)}$ around both points $z^{(b)}$ and $z^{(c)}$. The
phase of the complex function $f(z^{(a)})$ will change by $2\pi W_{a,bc}$
where $W_{a,bc}$ is an integer (see Fig. \ref{DDD}).  Since $f(z^{(a)})$
has an order $D_{ab}$ zero at $z^{(b)}$ and an order $D_{ac}$ zero at
$z^{(c)}$, the integer $W_{a,bc}$ satisfy
\begin{equation*}
 W_{a,bc} \geq D_{ab}+D_{ac} .
\end{equation*}
because $f(z^{(a)})$ may also have off-particle zeros.  Now let $z^{(b)}\to
z^{(c)}$ to fuse into $z^{(b+c)}$.  In this limit $W_{a,bc}$ becomes the order
of zeros between $z^{(a)}$ and $z^{(b+c)}$: $W_{a,bc}=D_{a,b+c}$. Thus we
obtain the following condition on $D_{ab}$: $ D_{a,b+c} \geq D_{ab}+D_{ac}$,
which gives us the second  concave condition \eq{Del3}.

We like to point out that the $n$-cluster condition has a very simple meaning in
the derived polynomial: $f(z^{(a)})$ has no off-particle zeros if $a=0$ mod
$n$.  So \ml{D_{a+b,n} = D_{a,n}+D_{b,n}} which leads to the cluster condition
\eq{clusterC}.

\subsubsection{Some additional conditions}

The two concave conditions are the main conditions on $\{S_a\}$.
We also have another condition
\ma{
\Del_2(a,a)=\text{even}
}
since the polynomial is a symmetric polynomial.
It turns out that we need yet another
a condition 
\ma{
\Del_3(a,b,c) = \text{even} .
}
It is hard to prove this mysterious condition using elementary methods.  Using
the connection between the symmetry polynomial and CFT (or vertex algebra), we
find that the condition $\Del_3(a,b,c) = \text{even}$ is directly related to
the requirement that the fermionic operators have half-integer scaling
dimensions and bosonic operators have integer scaling dimensions.\cite{LWW1024}

We conjecture that \emph{the patterns of zeros
\ml{(\frac{m}{n};S_2,\cdots,S_n)} that satisfy the above conditions describe
infinite-variable symmetric polynomials.}\cite{WW0808} Those
\ml{(\frac mn ;S_2,\cdots,S_n)} ``classify'' infinite-variable symmetric polynomials
and FQH states with filling fraction \ml{\nu=n/m}.

\subsubsection{Primitive solutions for pattern of zeros}

Let us list some patterns of zeros, \ml{(\frac{m}{n};S_2,\cdots,S_n)}, that
satisfy the above conditions.  We note that the conditions are semi-linear in
\ml{(\frac{m}{n};S_2,\cdots,S_n)}. So, if \ml{(\frac{m}{n}; S_2,\cdots,S_n)}
and \ml{(\frac{m'}{n'}; S_2',\cdots,S_n')}  are solutions, then \ml{
(\frac{m''}{n''}; S_2'',\cdots,S_n'') =(\frac{m}{n}; S_2,\cdots,S_n)
+(\frac{m'}{n'}; S_2',\cdots,S_n')} is also a solution.  Such a result has the
following meaning: Let $P(\{z_i\})$, $P'(\{z_i\})$, and $P''(\{z_i\})$ are
three symmetric polynomials described by pattern of zeros
\ml{(\frac{m}{n};S_2,\cdots,S_n)}, \ml{(\frac{m'}{n'};S_2',\cdots,S_n')}, and
\ml{(\frac{m''}{n''};S_2'',\cdots,S_n'')} respectively, we then have
\ml{P''(\{z_i\})=P(\{z_i\}) P'(\{z_i\})}.  Such a property allow us to
introduce the notion of primitive pattern of zeros as the patterns
of zeros that cannot to written as the sum of two other patterns of zeros.  In
this section, we will only list the primitive patterns of zeros.

\ml{1}-cluster state: \ml{\nu=1/k} Laughlin state
\ma{
P_{1/k}: \ \ \ \ \ \ \ \ (\frac mn;) &=(\frac{k}{1}; ),
\nonumber\\
(n_0,\cdots,n_{k-1})&=(1,0, \cdots,0) .
}

\ml{2}-cluster state: Pfaffian state (\ml{Z_2} parafermion state)
\ma{
P_{\frac{2}{2};Z_2}:\ \ \ \ (\frac mn;S_2) &=(\frac{2}{2};0),
\nonumber\\
(n_0,\cdots,n_{m-1})&=(2,0)
}

\ml{3}-cluster state: \ml{Z_3} parafermion state
\ma{
P_{\frac{3}{2};Z_3}:\ \ \ \ (\frac mn;S_2,S_3) &=(\frac{2}{3};0,0),
\nonumber\\
(n_0,\cdots,n_{m-1})&=(3,0)
}

\ml{4}-cluster state: \ml{Z_4} parafermion state
\ma{
P_{\frac{4}{2};Z_4}: (\frac mn;S_2,\cdots,S_4) &=(\frac{2}{4};0,0,0),
\nonumber\\
(n_0,\cdots,n_{m-1})&=(4,0),
}

\ml{5}-cluster states (we have two of them): \ml{Z_5} (generalized) parafermion states
\ma{
P_{\frac{5}{2};Z_5}: (\frac mn;S_2,\cdots,S_5) &= (\frac{2}{5};0,0,0,0),
\nonumber\\
(n_0,\cdots,n_{m-1})&=(5,0)
}
\ma{
P_{\frac{5}{8};Z_5^{(2)}}: (\frac mn;S_2,\cdots,S_5) &= (\frac{8}{5};0,2,6,10),
\nonumber\\
(n_0,\cdots,n_{m-1})&=(2,0,1,0,2,0,0,0)
}

\ml{6}-cluster state:
\ma{
P_{\frac{6}{2};Z_6}: (\frac mn;S_2,\cdots,S_6) &= (\frac{2}{6};0,0,0,0,0),
\nonumber\\
(n_0,\cdots,n_{m-1})&=(6,0)
}

\ml{7}-cluster states (we have four of them):
\ma{
P_{\frac{7}{2};Z_7}: (\frac mn;S_2,\cdots,S_7) &= (\frac{2}{7};0,0,0,0,0,0),
\nonumber\\
(n_0,\cdots,n_{m-1})&=(7,0)
}
\ma{
P_{\frac{7}{8};Z^{(2)}_7}: (\frac mn;S_2,\cdots,S_7) &= (\frac{8}{7};0,0,2,6,10,14),
\nonumber\\
(n_0,\cdots,n_{m-1})&=(3,0,1,0,3,0,0,0)
}
\ma{
P_{\frac{7}{18};Z^{(3)}_7}: (\frac mn;S_2,\cdots,S_7) &= (\frac{18}{7};0,4,10,18,30,42),
\nonumber\\
(n_0,\cdots,n_{m-1})=(2,&\,0,0,0,0,1,0,0,0,2,0,0,0,0,0)
}
\ma{
P_{\frac{7}{14};C_7}: (\frac mn;S_2,\cdots,S_7) &= (\frac{14}{7};0,2,6,12,20,28),
\nonumber\\
(n_0,\cdots,n_{m-1})=(2,&\,0,1,0,1,0,1,0,2,0,0,0,0,0)
}


\subsubsection{How good is the pattern-of-zeros classification?}

How good is the pattern-of-zeros classification?
Not so good, and not so bad.

Clearly, every symmetric polynomial \ml{P} corresponds to a unique pattern of
zeros \ml{\{S_a\}}. But only some patterns of zeros correspond to a unique
symmetric polynomial. So the pattern-of-zeros classification is not so good.  It
appears that all the primitive pattern of zeros correspond to a unique a unique
symmetric polynomial. Therefore, the pattern-of-zeros classification is not so
bad.

We also know that some composite patterns of zeros correspond a unique
symmetric polynomial, while other composite patterns of zeros do not correspond
a unique symmetric polynomial.  Let $P_{n_i}$ be a symmetric polynomial
described by a primitive pattern of zeros with an $n_i$-cluster.  It appear
that $P=\prod_i P_{n_i}$ will have a pattern of zeros that  corresponds a
unique symmetric polynomial if $n_i$'s has no common factor.

So only for certain patterns of zeros, the data
\ml{\{\frac{m}{n};S_2,...,S_n\}} contain all the information to fix the
symmetric polynomials.  In general, we need more information than
\ml{\{\frac{m}{n};S_2,...,S_n\}} to fully characterize symmetry polynomials of
infinite variables.

\subsection{Topological properties from pattern of zeros}

For those patterns of zeros that uniquely characterize the symmetry polynomials
of infinite variables (or FQH wave functions), we should be able to calculate 
the universal properties of the FQH states from the data 
\ml{(\frac{m}{n};S_2,\cdots,S_n)}.  Those universal properties include:
\begin{itemize}
\item
The filling fraction $\nu$.
\item
 Topological degeneracy on torus and other Riemann
surfaces
\item
 Number of quasiparticle types
\item
 Quasiparticle charges
\item
Quasiparticle scaling dimensions
\item
 Quasiparticle fusion algebra
\item
Quasiparticle statistics (Abelian and non-Abelian)
\item
 The counting of edge excitations (central charge
\ml{c} and spectrum)
\end{itemize}

At moment, we can calculate many of the above universal properties from the
pattern-of-zeros data \ml{(\frac{m}{n};S_2,\cdots,S_n)}.  For example, the
filling fraction $\nu$ is given by \ml{\nu=n/m}.  But we still do not know how
to calculate scaling dimensions and statistics for some of the quasiparticles.

In this subsection, we develop a pattern-of-zeros description
of the quasiparticle excitations in FQH states.
This will allow us to calculate many universal properties
from the pattern of zeros.

%

\subsubsection{Pattern of zeros of quasiparticle excitations}

 A quasiparticle is a defect in the ground state wave function \ml{P(\{z_i\})}.
It is a place where we have more power of zeros.  For example, the ground state
wave function of $\nu=1/2$ Laughlin state is given by \ml{\prod_{i<j}
(z_i-z_j)^2}.  The state with a quasiparticle at $\xi$ is given by \ml{\prod_i
(z_i-\xi) \prod_{i<j} (z_i-z_j)^2} (see section \ref{qpsec}).  
As we bring several $z_i$'s to $\xi$, \ml{\prod_i
(z_i-\xi) \prod_{i<j} (z_i-z_j)^2} vanishes according to a pattern of zeros.
In general, each
quasiparticle labeled by \ml{\ga} in a FQH state can be quantitatively
characterized by distinct pattern of zeros. 

\begin{figure}[tb]
\centerline{
\includegraphics[scale=0.7]{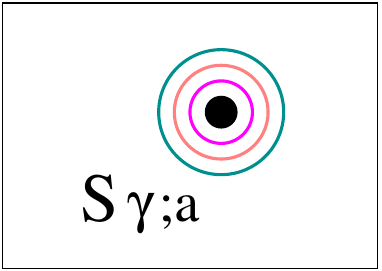}
}
\caption{
\sffamily
The graphic picture of the pattern of zeros
for a quasiparticle.
}
\label{qp}
\end{figure}

Let \ml{P_\ga(\xi;\{z_i\})} be the wave function
with a quasiparticle $\ga$ at \ml{z=\xi}.
To describe the structure of the zeros as we
bring  bosons to the quasiparticle,
we set \ml{z_i=\la \eta_i + \xi}, \ml{i=1,2,\cdots,a} 
and let $\la \to 0$:
\ma{
P_\ga(\xi;\{z_i\})=\la^{S_{\ga;a}} \t P_\ga(z^{(a)}=\xi, z_{a+1},z_{a+2},\cdots) + O(\la^{S_a+1})
}
\ml{S_{\ga;a}} is the order of zeros of \ml{P_\ga(\xi;z_i)}
when we bring \ml{a} bosons to  \ml{\xi}.
\emph{The sequence of integers \ml{\{S_{\ga;a}\}} is the quasiparticle
pattern of zeros that characterizes the quasiparticle \ml{\ga}.
}
We note that the ground-state pattern of zeros
 \ml{\{S_{a}\}} correspond to the trivial quasiparticle
\ml{\ga=0}: \ml{\{S_{0;a}\}=\{S_{a}\} }

 To find the allowed quasiparticles, we simply need to find
(i) the conditions that \ml{S_{\ga;a}} must satisfy and 
(ii) all the \ml{S_{\ga;a}} that satisfy those conditions.

\subsubsection{Conditions on quasiparticle
pattern of zeros $S_{\ga;a}$}

The quasiparticle
pattern of zeros also satisfy two
concave conditions
\ma{
& S_{\ga;a+b}- S_{\ga;a} - S_b \geq 0,
\\
&S_{\ga;a+b+c}
- S_{\ga;a+b}
- S_{\ga;a+c}
- S_{b+c}
+S_{\ga;a}
+S_{b}
+S_{c} \geq 0
}
and a cluster condition
\ma{
S_{\ga;a+kn}
=S_{\ga;a} + k(S_{\ga;n} + ma )+mn\frac{k(k-1)}{2}
}
The cluster condition implies that a finite sequence
\ml{(S_{\ga;1},\cdots,S_{\ga;n})} determines the infinity sequence
\ml{\{S_{\ga;a}\}}.

We can also use the sequence $l_{\ga;a}=S_{\ga,a}-S_{\ga,a-1}$ or
$n_{\ga;l}=\sum_{i=1} \del_{l,l_{\ga;i}}$ to describe the quasiparticle
sequence $S_{\ga;a}$.  The $n_{\ga;l}$ description is simpler and reveals
physical picture more clearly than $S_{\ga;a}$.

\subsubsection{The solutions for the quasiparticle patterns of zeros}

We can find all \ml{(S_{\ga;1},\cdots,S_{\ga;n})} that satisfy the above
concave and cluster conditions through numerical calculations. This allow us to
obtain all the quasiparticles.

For the \ml{\nu=1} Pfaffian state (\ml{n=2} and \ml{m=2})
described by
\ma{
S_1,S_2,\cdots: &\ 0,0, 2, 4, 8, 12, 18, 24, \cdots
\nonumber\\
n_0n_1n_2\cdots:&\ 2020202020202020202\cdots ,
}
we find that the quasiparticle patterns of zeros are given by
(expressed in terms of $n_{\ga,l}$)
\ma{
n_{\ga;0} n_{\ga;1} n_{\ga;2}\cdots:&\ 2020202020202020202\cdots
\ \ \ \ Q_\ga=0
\nonumber\\
n_{\ga;0} n_{\ga;1} n_{\ga;2}\cdots:&\ 0202020202020202020\cdots
\ \ \ \ Q_\ga=1
\nonumber\\
n_{\ga;0} n_{\ga;1} n_{\ga;2}\cdots:&\ 1111111111111111111\cdots
\ \ \ \ Q_\ga=1/2
}

The above three pattern of zeros are not all the solutions of the quasiparticle
conditions.  However, all other quasiparticle solutions can be obtained from
the above three by removing some bosons.  Those quasiparticle solutions are
equivalent to one of the above three solutions.  For example $n_{\ga;0}
n_{\ga;1} \cdots=102020202\cdots$, $n_{\ga;0} n_{\ga;1}
\cdots=002020202\cdots$, \etc are also quasiparticle solutions which are
equivalent to $n_{\ga;0} n_{\ga;1} \cdots =202020202\cdots$.  Therefore, we
find that the \ml{\nu=1} Pfaffian state has three types of quasiparticles.

We note that the ground state degeneracy on torus is equal to the number of
quasiparticle types.  So the \ml{\nu=1} Pfaffian state has a three-fold
degeneracy on a torus.  The charge of quasiparticles can be also calculated
from the quasiparticle pattern of zeros:
\ma{
Q_\ga=\frac{1}{m} \sum_{a=1}^n (l_{\ga;a}-l_a)
=\frac 1m (S_{\ga;n} -S_{n} ).
}

%
%

Let us list the
number of quasiparticle types calculated from pattern of zeros
for various FQH states.
For the parafermion states \ml{P_{\nu=\frac{n}{2};Z_n}} (\ml{m=2}),
\begin{align*}
\begin{tabular}{|c|c|c|c|c|c|c|c|c|}
\hline
$P_{\frac{2}{2};Z_2}$ &
$P_{\frac{3}{2};Z_3}$ &
$P_{\frac{4}{2};Z_4}$ &
$P_{\frac{5}{2};Z_5}$ &
$P_{\frac{6}{2};Z_6}$ &
$P_{\frac{7}{2};Z_7}$ &
$P_{\frac{8}{2};Z_8}$ &
$P_{\frac{9}{2};Z_9}$ &
$P_{\frac{10}{2};Z_{10}}$
\\
\hline
3  &
4  &
5  &
6  &
7  &
8  &
9  &
10  &
11
\\
\hline
 \end{tabular}
\end{align*}

For the parafermion states \ml{P_{\nu=\frac{n}{2+2n};Z_n}} (\ml{m=2+2n})
\begin{align*}
 \begin{tabular}{|c|c|c|c|c|c|c|c|c|}
\hline
$P_{\frac{2}{6};Z_2}$ &
$P_{\frac{3}{8};Z_3}$ &
$P_{\frac{4}{10};Z_4}$ &
$P_{\frac{5}{12};Z_5}$ &
$P_{\frac{6}{14};Z_6}$ &
$P_{\frac{7}{16};Z_7}$ &
$P_{\frac{8}{18};Z_8}$ &
$P_{\frac{9}{20};Z_9}$ &
$P_{\frac{10}{22};Z_{10}}$
\\
\hline
9  &
16  &
25  &
36  &
49  &
64  &
81  &
100  &
121
\\
\hline
 \end{tabular}
\end{align*}

For the generalized parafermion states \ml{P_{\nu=\frac{n}{m};Z_n^{(k)}}}
\begin{align*}
 \begin{tabular}{|c|c|c|c|c|c|c|c|}
\hline
$P_{\frac{5}{8};Z_5^{(2)}}$ &
$P_{\frac{5}{18};Z_5^{(2)}}$ &
$P_{\frac{7}{8};Z_7^{(2)}}$ &
$P_{\frac{7}{22};Z_7^{(2)}}$ &
$P_{\frac{7}{18};Z_7^{(3)}}$ &
$P_{\frac{7}{32};Z_7^{(3)}}$ &
$P_{\frac{8}{18};Z_8^{(3)}}$ &
$P_{\frac{9}{8};Z_9^{(2)}}$
\\
\hline
24  &
54  &
32  &
88  &
72  &
128  &
81  &
40
\\
\hline
 \end{tabular}
\end{align*}
where \ml{k} and \ml{n} are co-prime.

For the composite parafermion states
\ml{
P_{\frac{n_1}{m_1};Z_{n_1}^{(k_2)}}
P_{\frac{n_2}{m_2};Z_{n_2}^{(k_2)}}
} obtained as products of two parafermion wave functions
\begin{align*}
 \begin{tabular}{|c|c|c|c|}
\hline
$ P_{\frac{2}{2};Z_2} P_{\frac{3}{2};Z_3} $ &
$ P_{\frac{3}{2};Z_3} P_{\frac{4}{2};Z_4} $ &
$ P_{\frac{2}{2};Z_2} P_{\frac{5}{2};Z_5} $ &
$ P_{\frac{2}{2};Z_2} P_{\frac{5}{8};Z_5^{(2)}} $
\\
\hline
30  &
70  &
63  &
117
\\
\hline
 \end{tabular}
\end{align*}
where \ml{n_1} and \ml{n_2} are co-prime.  The inverse
filling fractions of the above
composite states are \ml{\frac{1}{\nu}=
\frac{1}{\nu_1}+\frac{1}{\nu_2}
=\frac{m_1}{n_1}+\frac{m_2}{n_2}
}.
More results can be found in \Ref{WW0809}.

All those results from the pattern of zeros agree
with the results from parafermion CFT: \cite{BW0932}
\ma{
\# \text{ of quasiparticles} = \frac{1}{\nu}  \prod_i \frac{n_i(n_i+1)}{2}
}
for the generalized composite parafermion state
\ma{
P &=\prod_i P_{\frac{n_i}{m_i};Z^{(k_i)}_{n_i}},
\ \ \
\{n_i\} \text{ co-prime},
\ \ \
(k_i,n_i) \text{ co-prime}.
}
The filling fraction for such generalized composite parafermion state is given
by $\nu =\Big( \sum_i \frac{m_i}{n_i}\Big)^{-1}$.

\subsubsection{Quasiparticle fusion algebra: $\ga_1 \ga_2=\sum_{\ga_3}
N^{\ga_3}_{\ga_1\ga_2} \ga_3 $
}

\begin{figure}[tb]
\centerline{
\includegraphics[scale=0.60]{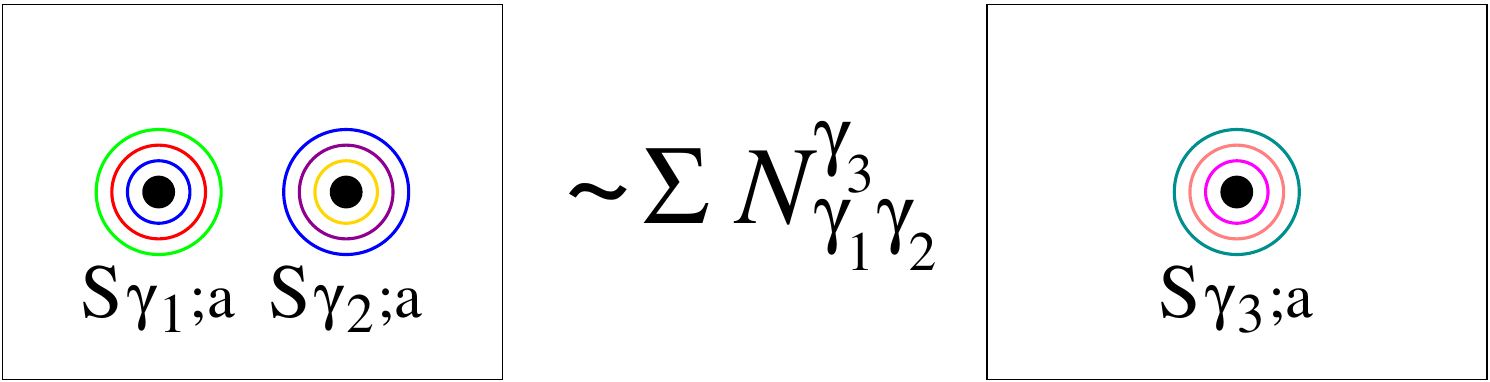}
}
\caption{
\sffamily
The graphic picture of the fusion of two quasiparticles.  Each box represent a
many-boson wave function.  In the left box, we have quasiparticle $\ga_1$ and
$\ga_2$ described by patterns of zeros $S_{\ga_1;a}$ and $S_{\ga_2;a}$.  Far
away from the two  quasiparticles, the wave function may contain several
different patterns of zeros $S_{\ga_3;a}$ that correspond to several different
quasiparticle types $\ga_3$.  So we say that $\ga_1$ and $\ga_2$ may fuse into
several different types of quasiparticles labeled by  $\ga_3$.
}
\label{qpfuse}
\end{figure}

When we fuse quasiparticles $\ga_1$ and $\ga_2$ together, we can get a third
quasiparticle $\ga_3$.  However, for non-Abelian quasiparticles, the fusion can
be more complicated.  Fusing $\ga_1$ and $\ga_2$ may produce several kind of
quasiparticles. Such kind of fusion is described by quasiparticle fusion
algebra (see Fig. \ref{qpfuse}): $\ga_1 \ga_2=\sum_{\ga_3}
N^{\ga_3}_{\ga_1\ga_2} \ga_3$, where $N^{\ga_3}_{\ga_1\ga_2}$ are non-negative
integers.

To calculate the fusion coefficients $N^{\ga_3}_{\ga_1\ga_2}$
from the pattern of zeros, let us put the
quasiparticle $\ga_1$ at $z=0$.  Far away from $z=0$, such a quasiparticle has
a pattern of zeros $n_{\ga_1;l}$ (in the occupation representation).  We then
insert a quasiparticle $\ga_2$ at $z=R$ for a large $R$.  At $z=r\gg R$, the
occupation becomes the occupation of the quasiparticle $\ga_3$:
$n_{\ga_3;l}$.
We see the the fusion of $\ga_2$ changes the
occupation pattern from   $n_{\ga_1;l}$ to $n_{\ga_3;l}$:
\ma{ &n_{\ga_1;0} n_{\ga_1;1} \cdots n_{\ga_1;a} [\ga_2]
n_{\ga_3;a+1} n_{\ga_3;a+2} \cdots
\nonumber \\
&\includegraphics[scale=0.65]{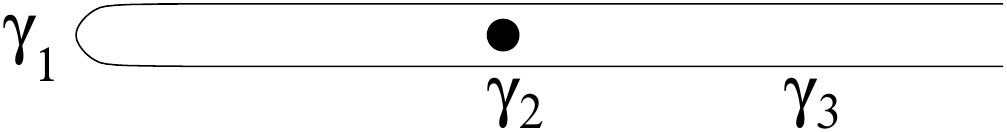}
}
So the the quasiparticle $\ga_2$ becomes a  ``domain wall'' between the $\ga_1$
occupation pattern and the $\ga_3$ occupation pattern.\cite{ABK0816}

From the above domain wall structure, we
can see only \ml{n_{\ga_1;l}} and \ml{n_{\ga_3;l}}, but we cannot see
\ml{n_{\ga_2;l}}. But this is enough for us.
We are able to find a condition on
\ml{n_{\ga_2;l}} so that it can induce a domain wall
between \ml{n_{\ga_1;l}} and \ml{n_{\ga_3;l}}: \cite{BW0932}
\ma{
\sum_{j=1}^{b} \Big(
l^\text{sc}_{\gamma_1;j+a}  
+l^\text{sc}_{\gamma_2;j+c}
\Big)
\leq  \sum_{j=1}^{b} \Big(
l^\text{sc}_{\gamma_3;j+a+c} 
+  l^\text{sc}_j
\Big)
}
for any \ml{a,b,c \in Z_+},
where \ml{l^\text{sc}_{\ga;a}=l_{\ga;a} -\frac{m(Q_\ga+a-1)}{n}}.

Solving the above equation allows us to determine when
\ml{N^{\ga_3}_{\ga_1\ga_2}} can be non-zero.  If we further assume that
\ml{N^{\ga_3}_{\ga_1\ga_2}=0,1}, then the fusion algebra can be determined.
Knowing $N^{\ga_3}_{\ga_1\ga_2}$ allows us to determine the ground state
degeneracies of FQH state on any closed Riemann surfaces.

We like to mention that for the generalized composite parafermion states which
have a CFT description, the pattern-of-zeros approach and the CFT approach give
rise to the same fusion algebra.  However, the pattern-of-zeros approach
applies to other FQH states whose CFT may not be known.

\section{The Vertex-algebra+pattern-of-zeros approach}


\subsection{\ml{Z}-graded vertex algebra}

The symmetric polynomial $P(\{z_i\})$
and the corresponding derived 
 polynomial $P_\text{derived}(\{z_i^{(a_i)}\})$
can be expressed as correlation functions
in a vertex algebra:
\ma{
 P(\{z_i\}) &= \<\prod_{i} V(z_i) \>, \ \ \ \
%
 P_\text{derived}
(\{z_i^{(a)}\}) = \<\prod_{i,a} V_a(z_i^{(a)}) \>
\nonumber\\
V_a(z) & = V^a
,\ \ \ \ V_a V_b=V_{a+b}.
}
The vertex algebra is generated by vertex operator $V(z)$
and is described by the following operator product expansion:
\ma{
V_a(z)V_b(w) =\frac{C_{ab}}{(z-w)^{h_a+h_b-h_{a+b}}} V_{a+b}(w)
+ ...
}
where $h_a$ is the scaling dimension of $V_a$ and $C_{ab}$ the structure
constant of the vertex algebra.  Such a vertex algebra is a \ml{Z}-graded
vertex algebra.

The pattern of zeros $S_a$ discuss before
is directly related to $h_a$:
\ma{
h_{a+b} -h_a-h_b=D_{a,b}=S_{a+b}-S_a-S_b
}
The \ml{n}-cluster condition implies that \ml{h_{a}\propto a^2}
if \ml{a =0 \text{ mod } n}.
This allows us to obtain
\ma{
h_a &= S_a- \frac{aS_n}{n}+\frac{am}{2}
}
We see that the pattern of zeros $S_a$ only describe the scaling dimensions of
the vertex operators. It does not describe the structure constants $C_{a,b}$.
So a more complete characterization of FQH wave functions (symmetric
polynomials) is given by \ml{(\frac{m}{n};S_a;C_{ab},...)}.  But
\ml{(\frac{m}{n};S_a;C_{ab},...)} may be an overkill.  We like to find out what
is the minimal set of date that can completely characterize the  FQH wave
functions (or the symmetric polynomials).

\subsection{$Z_n$-vertex algebra}

If the above \ml{Z}-graded vertex algebra 
satisfies the $n$-cluster condition, then it can be viewed  a
\ml{Z_n}-vertex algebra $\otimes$ a \ml{U(1)} current
algebra:
\ma{
V_a(z)=\psi_a(z) e^{i a \phi(z)\sqrt{m/n}}
}
where \ml{j=\prt \phi} generates the \ml{U(1)} current
algebra and \ml{\psi_a} generates the \ml{Z_n}-vertex
algebra:
\ma{
\psi_a(z)\psi_b(w) &=\frac{C_{ab}}
{(z-w)^{h^\text{sc}_a+h^\text{sc}_b-h^\text{sc}_{a+b}}} \psi_{a+b}(w)
+ ...
}
where
$\psi_n =1$ as the result of the $n$-cluster condition.
The scaling dimension of \ml{\psi_a(z)} is
\ma{
h^\text{sc}_a &= h_a-\frac{a^2m}{2n}
=S_a- \frac{aS_n}{n}+\frac{am}{2}-\frac{a^2m}{2n},
&
h^\text{sc}_a&=h^\text{sc}_{a+n}
}
The two sets of data \ml{(\frac{m}{n};S_2,...,S_n)} and
\ml{(\frac{m}{n};h^\text{sc}_1,...,h^\text{sc}_{n-1})} completely determine
each other:
\begin{align}
 S_a=h^\text{sc}_a-a h^\text{sc}_1+\frac{a(a-1)m}{2n}.
\end{align}
So we can also use \ml{(\frac{m}{n};h^\text{sc}_1,...,h^\text{sc}_{n-1})}
to describe the pattern of zeros.

From the pattern-of-zeros consideration, 
we find that $h^\text{sc}_a$ must satisfy
\ma{
&S_a = h^\text{sc}_a-ah^\text{sc}_1+\frac{a(a-1)m}{2n}=
\text{integer} \geq 0
\nonumber\\
&h^\text{sc}_{a+b}-h^\text{sc}_a-h^\text{sc}_b + \frac{abm}{n}
= D_{ab} =\text{integer} \geq 0
\\
&h^\text{sc}_{a+b+c} -h^\text{sc}_{a+b}
-h^\text{sc}_{b+c} -h^\text{sc}_{a+c} +h^\text{sc}_a +h^\text{sc}_b
+h^\text{sc}_c
\nonumber\\
&\ \ \ \ = \Del_3(a,b,c) = \text{even integer} \geq 0
}
But the above conditions are only on $h^\text{sc}_a$.  To get the conditions on
$C_{ab}$, we can use the generalized Jacobi identity\cite{N0774} to obtain a
set a non-linear equations for \ml{( h^\text{sc}_a, C_{ab},...)}.\cite{LWW1024}
Those conditions may be sufficient and necessary which may lead to a
classification of \ml{Z_n}-vertex algebra.  

For some simple pattern of zeros $h^\text{sc}_a$, we are able to build a closed
set of non-linear equations for \ml{( h^\text{sc}_a, C_{ab},...)}, which lead
to a well defined \ml{Z_n}-vertex algebra.  This allows us to calculate
quasiparticle scaling dimensions, quasiparticle statistics, central charge
(edge spectrum), ...\cite{LWW1024} We would like to point out that in
\Ref{S0802,FS1115}, a very interesting approach based the pattern of zeros and
modular transformation of torus is proposed, that allows us to calculate the
fractional statistics of some quasiparticles directly from the pattern-of-zeros
data.  We also like to point out that finding valid \ml{( h^\text{sc}_a,
C_{ab},...)} corresponds to finding a well defined $Z_n$ vertex algebra.
Finding the quasiparticle patterns of zeros corresponds to finding the
representations of the $Z_n$ vertex algebra.

But at moment, we cannot handle more general pattern of zeros $h^\text{sc}_a$,
in the sense that we have some difficulties to obtain a closed set of non-linear
algebraic equations for \ml{( h^\text{sc}_a, C_{ab},...)}.  We hope that, after
some further research, the pattern-of-zeros approach may lead to a
classification of $Z_n$-vertex algebra, which in turn lead to a classification
of symmetric polynomials and FQH states.

\section{Summary}

Although still incomplete, the pattern-of-zeros approach provides quite a
powerful way to study symmetric polynomials with infinite variables and FQH
states.  It connects several very different fields, such as strongly correlated
electron systems, topological quantum field theory, CFT (for the edge states),
modular tensor category theory (for the quasiparticle statistics), and a new
field of infinite-variable symmetric polynomial.  This article only reviews the
first step in this very exciting direction.  More exciting results are yet to
come.

This research is supported by NSF Grant No. DMR-1005541 and NSFC 11074140.

\bibliography{../../bib/wencross,../../bib/all,../../bib/publst}

\end{document}